\documentclass[usenatbib]{mn2e}
\usepackage[dvips]{graphicx}
\def\cn2{\mbox{$C_N^2$}}

\title[Optical turbulence simulations at Mt Graham]{Optical turbulence simulations at Mt Graham using the Meso-NH model}
\author[S. Hagelin et al.]{S. Hagelin,$^{1, 2}$ \thanks{E-mail:
   hagelin@arcetri.astro.it; masciadri@arcetri.astro.it},
 E. Masciadri$^1$\footnotemark[1], F. Lascaux$^1$ \\ 
 $^1$INAF Osservatorio Astrofisico di Arcetri, Largo Enrico Fermi 5, I-501 25 Florence, Italy \\ 
 $^2$Uppsala Universitet, Department of Earth Sciences, Villav\"agen 16, S-752 36 Uppsala, Sweden}

\begin{document}
\label{firstpage}
\date{Accepted 2010 November 25. Received 2010 November 12; in original form 2010 September 29}
\pagerange{\pageref{firstpage}--\pageref{lastpage}}
\pubyear{2010}

\maketitle

\begin{abstract}
The mesoscale model Meso-NH is used to simulate the optical turbulence at Mt Graham (Arizona, US), site of the Large Binocular Telescope. Measurements of the $\cn2$-profiles obtained with a Generalized Scidar from 41 nights are used to calibrate and quantify the model ability in reconstructing the optical turbulence above the site. The measurements are distributed over different periods of the year permitting us to study the model performance in different seasons. A statistical analysis of the simulations is performed for all the most important astroclimatic parameters: the $\cn2$-profiles, the seeing $\varepsilon$, the isoplanatic angle $\theta_{0}$ and the wavefront coherence time $\tau_{0}$. 

The model shows a general good ability in reconstructing the morphology of the optical turbulence (shape of the vertical distribution of the $\cn2$) as well as the strength of all the integrated astroclimatic parameters. The relative error (with respect to measurements) of the averaged seeing on the whole atmosphere for the whole sample of 41 nights is within 9.0\%. The median value of the relative error night by night is equal to 18.7\% so that the model still maintains very good performances. Comparable percentages are observed in partial vertical slabs (free atmosphere and boundary layer) and in different seasons (summer and winter). We prove that the most urgent problem, at present, is to increase the ability of the model in reconstructing very weak and very strong turbulence conditions in the high atmosphere. This evidence in the model mainly affects, at present, the model performances for the isoplanatic angle predictions for which the median value of the relative error night by night is equal to 35.1\%. No major problems are observed for the other astroclimatic parameters. A variant to the standard calibration method is tested but we find that it does not provide better results confirming the solid base of the standard method.

\end{abstract}


\section{Introduction}
Mt Graham International Observatory is located in south-eastern Arizona, USA, at 3200 m above sea level. The observatory consists of three telescopes: the Large Binocular Telescope (LBT - with two 8.4 m mirrors), the Heinrich Hertz Sub-millimetre Telescope (SMT - D=10 m) and the Vatican Advanced Technology Telescope (VATT - D=1.83 m). Some studies aiming to characterize the optical turbulence above Mt Graham have been done in the past with measurements mainly retrieved from a Generalized Scidar (GS) \citep{Egner07, Mas10}. 
In this paper we investigate the possibility to characterize and predict the optical turbulence (vertical distribution and integrated values) at Mt Graham using an atmospheric mesoscale model called Meso-NH. In a previous paper \citep{Hag10} we have used the same model to investigate the possibility to predict the vertical wind speed distribution at Mt Graham. We proved that the Meso-NH model provides reliable estimates of the vertical distribution of the wind speed at all heights from the ground up to 20 km. This wind speed can be used therefore for the calculation of the wavefront coherence time ($\tau_{0}$) on the Mt Graham summit.

The Meso-NH model has already been used to study the optical turbulence (OT) at different astronomical sites. It was first used by \cite{Mas99a,Mas99b} who also developed the code for the optical turbulence in the Meso-NH, the so called Astro-Meso-NH package including the algorithms for the $\cn2$ parametrization. These were, to our knowledge, the first $\cn2$-profiles ever simulated with a mesoscale model in an astronomical context. In those studies the model was proved to be sensitive to orographic effects and to be able to reconstruct $\cn2$-profiles well correlated to measurements provided by a Scidar. Also it was able to discriminate between the worst and the best seeing of the measurement campaign. \cite{Mas99a,Mas99b} discussed the main limitations encountered in the $\cn2$ modelling and proposed methods to overcome them.

Some improvements in the model reliability have been achieved more recently thanks to the method for the model calibration introduced by \cite{Mas01}. Such a calibration reduces some systematic errors and it is based on the tuning of a free parameter called minimum turbulence kinetic energy (E$_{min}$). The authors proved that the $\cn2$ can be related to E$_{min}$ and the calibration aims to fix the value of E$_{min}$ using as a reference the measured $\cn2$-profiles. \cite{Mas01} proved that the calibrated Meso-NH model reduced some of the previous systematic errors in the model and the shape of the resulting $\cn2$-profiles fitted better with measurements than the uncalibrated model. The statistic reliability of this method has been proved only later by \cite{Mas04} who used the calibrated Meso-NH to simulate the optical turbulence at the San Pedro M\'artir Observatory on a sample of 10 nights, for which there were $\cn2$-profiles obtained with a Generalized Scidar and micro-thermal sensors mounted on radiosondes. Measurements provided by different instruments permitted the authors to prove that the dispersion between measurements and simulations was of the same order as the dispersion between measurements provided by different instruments. This qualified the numerical technique and the Meso-NH model as potentially useful to perform autonomous estimates of the optical turbulence. More recently \citep{Mas06} the Meso-NH has been used in an autonomous way (after calibration) on a sample of 80 nights to simulate the optical turbulence at San Pedro M\'artir. 

The algorithms for the optical turbulence parametrization introduced by \cite{Mas99a} as well as the calibration procedure \citep{Mas01} have been later on implemented in other mesoscale models such as WRF \citep{Klemp08, Bus10}.

The goal of this paper is to use the Meso-NH model to simulate the optical turbulence at Mt Graham using, as a reference, a sample of measurements associated to 43 nights. A sub-sample of this rich statistical sample is used to calibrate the model. The outputs of the model are then compared to measurements obtained with a Generalized Scidar \citep{Mas10}. The sample of measurements is around four times larger than the previous samples \citep{Mas04} and the largest ever used at present for this purpose. For this reason we can: {\bf (1)} verify if the calibration permits the model to still obtain a typical vertical distribution well correlated with measurements and/or if this correlation increases/decreases; {\bf (2)} study how the correlation between numerical calculations and measurements is deteriorated when we consider in the sample data not used for the calibration; {\bf (3)} study the ability of the Meso-NH model in reproducing the seasonal differences in the optical turbulence. The measurements are, indeed, quite evenly distributed over different periods. In this paper we also test a variant of the method of model calibration proposed by \cite{Mas01}, hereafter MJ01.  

In Section \ref{sec:meas} we briefly present the measurements used as a reference in this study. In Section \ref{sec:mnh} we present the Meso-NH model and describe the model configuration used in this study. In Section \ref{sec:calib} we describe the calibration procedure. In Section \ref{sec:res} we present the results of this study in four subsection dedicated respectively to the vertical distribution of the optical turbulence ($\cn2$-profiles), the seeing $\varepsilon$, the isoplanatic angle $\theta_{0}$ and the wavefront coherence time $\tau_{0}$. In Section \ref{sec:concl}  the conclusions of this study are presented. 

\begin{figure*}
\includegraphics[width=5.5cm]{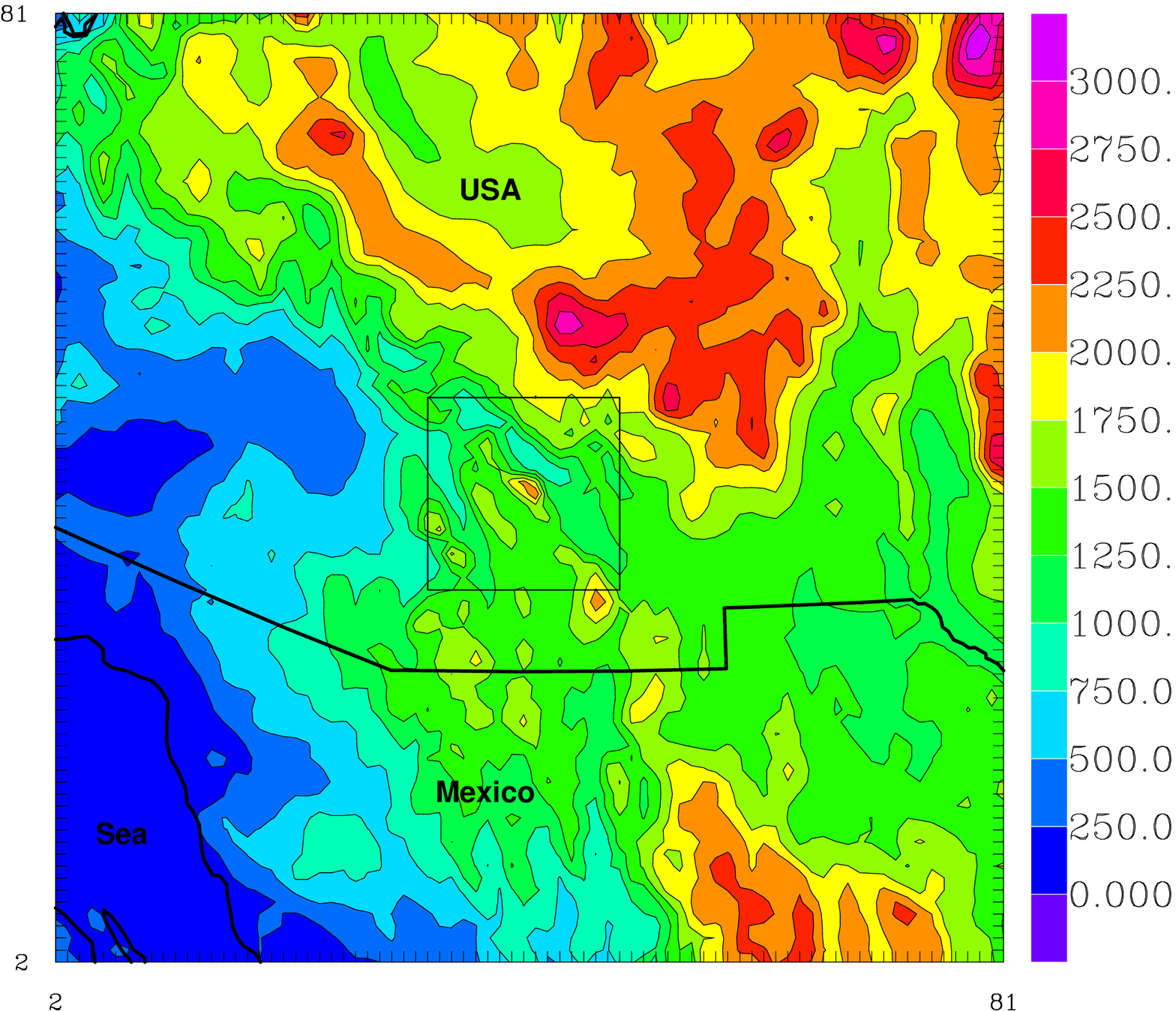}
\includegraphics[width=5.5cm]{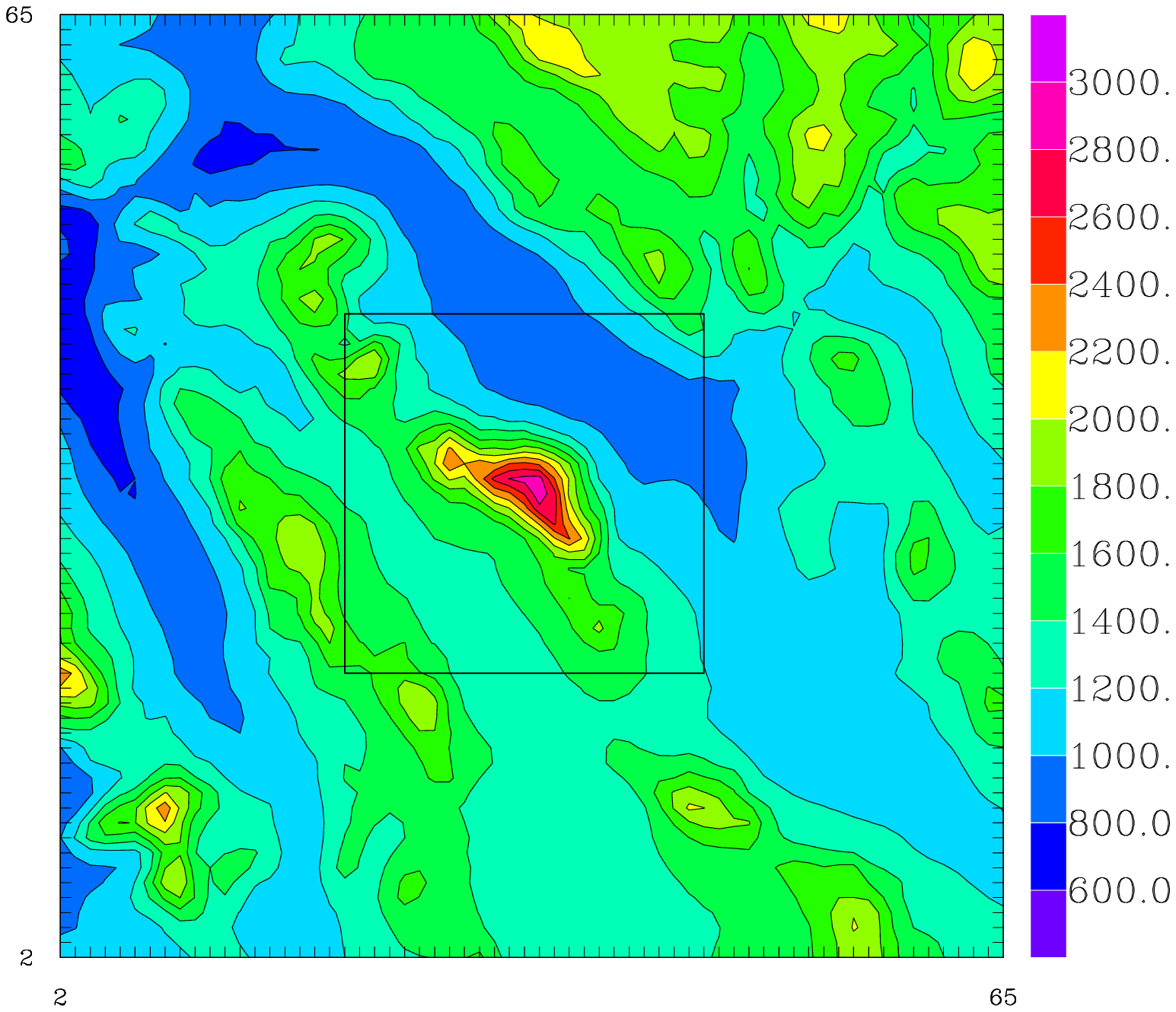}
\includegraphics[width=5.5cm]{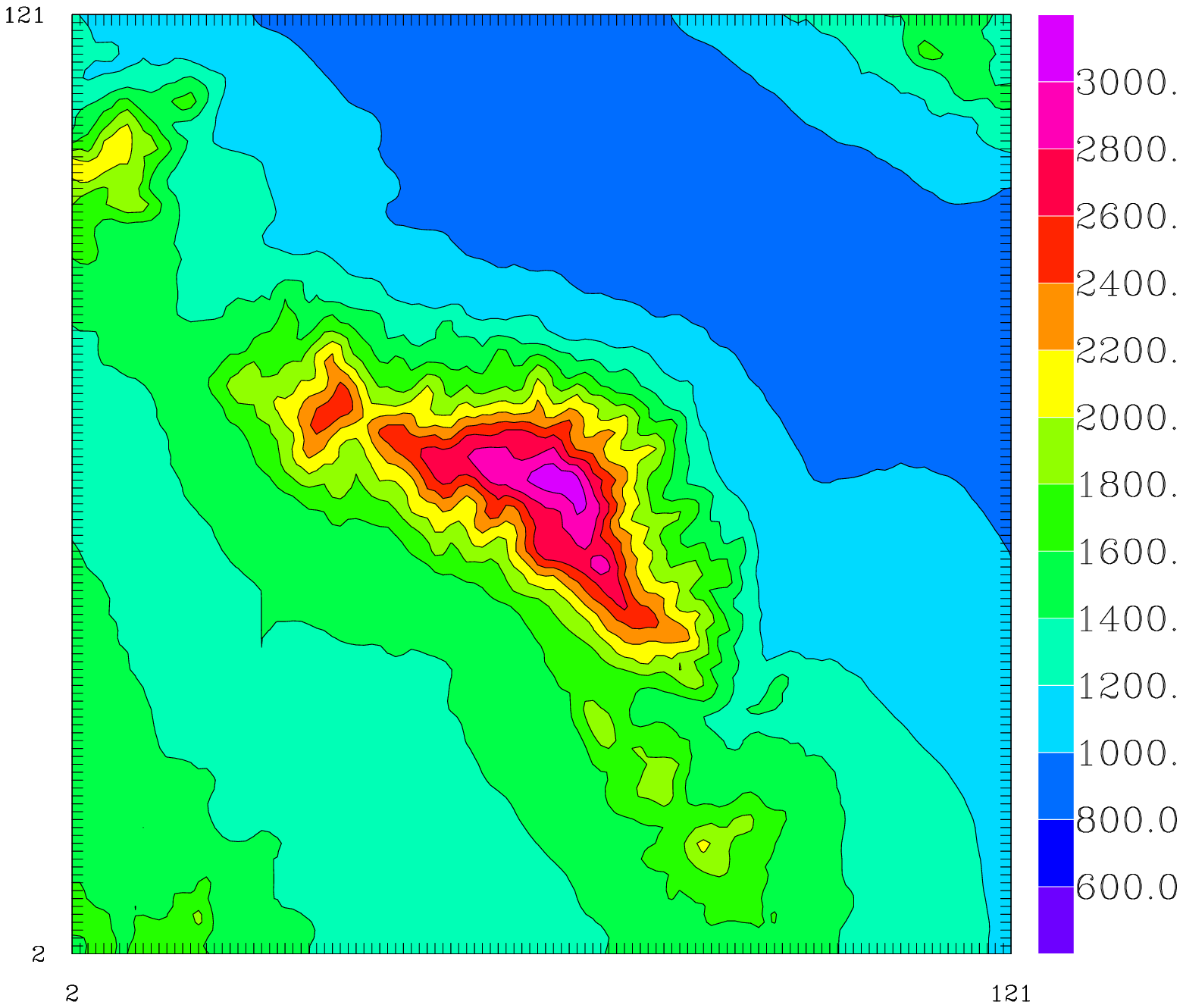}
\caption{Topographic areas covered by the models. The left hand panel is the outermost model (Model 1) with a resolution of 10 km. The black square indicates the location of the sub-model (Model 2). The central panel is the sub-model with 2.5 km resolution. The black square show the location of the innermost model. The right panel shows the innermost model (Model 3) with a 500 m resolution. The colour-scale indicates the height above sea level.}
\label{maps}
\end{figure*}


\section{Measurements of reference}
\label{sec:meas}
$\cn2$ profiles obtained with a Generalized Scidar (GS) \citep{Egner07, Mas10} mounted on the VATT have been used as a reference for this study. The GS is based on the observation of binaries with a typical separation $\theta$ within (3-10) arcsec, the binary magnitude m$_{1}$,m$_{2}$ $\le$ 5 mag and $\Delta$(m$_{1,2}$) $\le$ 1 mag. The GS needs a telescope with a pupil size $\ge$ 1.5 m. The $\cn2$-profiles are obtained from the inversion of the auto-correlation of the scintillation map of binaries \citep{Fuchs98,Avila97}. This instruments provides a vertical distribution of the optical turbulence on the whole 20 km from the ground with a vertical resolution that scales as $0.78\sqrt{\lambda (h - h_{gs})}$ / $\theta$ where $h$ is the height from the ground and h$_{gs}$ is the height under ground at which the conjugated plane is optically placed. The GS provides a vertical resolution that, considering the parameter space just described, is typically of the order of 1km on the whole atmosphere, reaching the best resolution near the ground (order of some hundreds of meters). GS measurements are the best choice (with respect to other vertical profilers) for a mesoscale model validation for a few reasons: (1) measurements are obtained with a remote sensing technique and they are therefore available for an extended period of time during a night, (2) the $\cn2$ is obtained with a completely independent and auto-consistent method requiring no calibration (3) one can access directly the $\cn2$ i.e. the prime element from which all the integrated astroclimatic parameters can be calculated autonomously. Appendix A reports the analytical equations that describe how the seeing, the isoplanatic angle and the wavefront coherence time are calculated from the $\cn2$.


\section{Model configuration}
\label{sec:mnh}
The Meso-NH is a non-hydrostatic mesoscale model developed jointly by the Centre National des Recherches M\'et\'eorologiques (CNRM - M\'et\'eo-France) and Laboratoire d'A\'erologie in Toulouse (France) \citep{LaF98}. It is a grid point model based on the anelastic approximation to efficiently filter out the acoustic waves. A \citet{gcs} coordinate on the vertical and a C-grid in the formulation of \citet{am} for the spatial digitalization is used. The temporal scheme is an explicit three-time-level leap-frog scheme with a time filter \citep{as}. The turbulent scheme is a one-dimensional 1.5 closure scheme \citep{cux}. The model permits the use of different mixing lengths. In this paper we used the 1-D \citet{bl} mixing length (BL89). The surface exchanges are computed in an externalized surface scheme (SURFEX) including the physical package ISBA (Interactions Soil Biosphere Atmosphere) \citep{np} that controls the air/ground turbulent fluxes budget of Meso-NH. The model can simulate the temporal evolution in three dimensions of the classical meteorological parameters, such as wind speed and direction, potential temperature, pressure, and so on. Meso-NH uses the code implemented by \cite{Mas99a,Mas99b} to forecast the optical turbulence ($C_N^2$ 3D maps) and all the astroclimatic parameters deduced from the $C_N^2$. We will refer to the 'Astro-Meso-NH code' to indicate this package. The integrated astroclimatic parameters are calculated integrating the  $C_N^2$ with respect to the zenith in the Astro-Meso-NH code.

The Meso-NH model is run, in this paper, in a grid nesting mode using three imbricated models with different resolutions (Table \ref{res}). They are all centred on the Mt Graham Observatory (32.7013${\degr}$N, 109.8919${\degr}$W). The size of the outermost model (Model 1) is 800x800 km which covers most part of south-eastern Arizona, south-west New Mexico and also a part of north-western Mexico, see Fig.~\ref{maps}. The middle model (Model 2) covers 160x160 km, using a grid-size of 2.5 km. The innermost model (Model 3) has a resolution of 500 m and is covering an area of 60x60 km. The vertical resolution is the same for all three models with 54 vertical grid points, reaching up to 20 km above the ground. The first vertical grid point is located 20 m above the ground and thereafter the grid is determined by a logarithmic stretching (20$\%$) up to 3.5 km above the ground. Above 3.5 km the resolution is almost constant and equal to $\sim$600 m.

\begin{table}
\caption{Meso-NH configuration. The second column gives the horizontal resolution $\Delta$X of the imbricated models, the third column the grid-points, the fourth column the surface covered by the models and the fifth column the time steps used by the model.}
\begin{tabular}{ccccc}
\hline
& $\Delta$X  & Grid points & Surface & Time step \\
& (km) &  & (km) & (s)\\
\hline
Model 1 & 10  & 80 x 80 & 800 x 800 & 30\\
Model 2 & 2.5 & 64 x 64 & 160 x 160 & 6\\
Model 3 & 0.5 & 120 x 120 & 60 x 60 & 3\\
\hline
\end{tabular}
\label{res}
\end{table}

\begin{figure*}
\includegraphics[width=7.5cm]{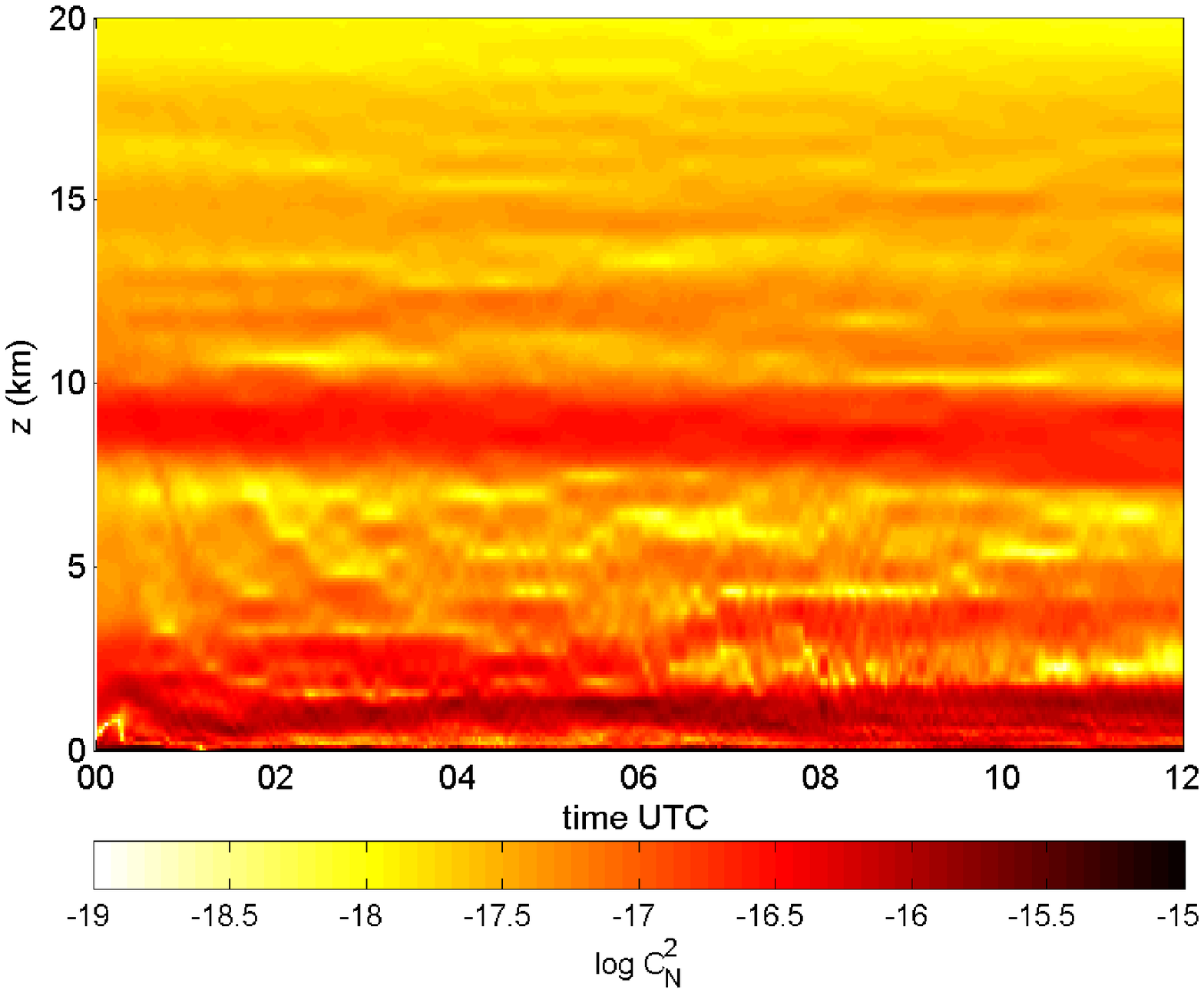}
\includegraphics[width=7.5cm]{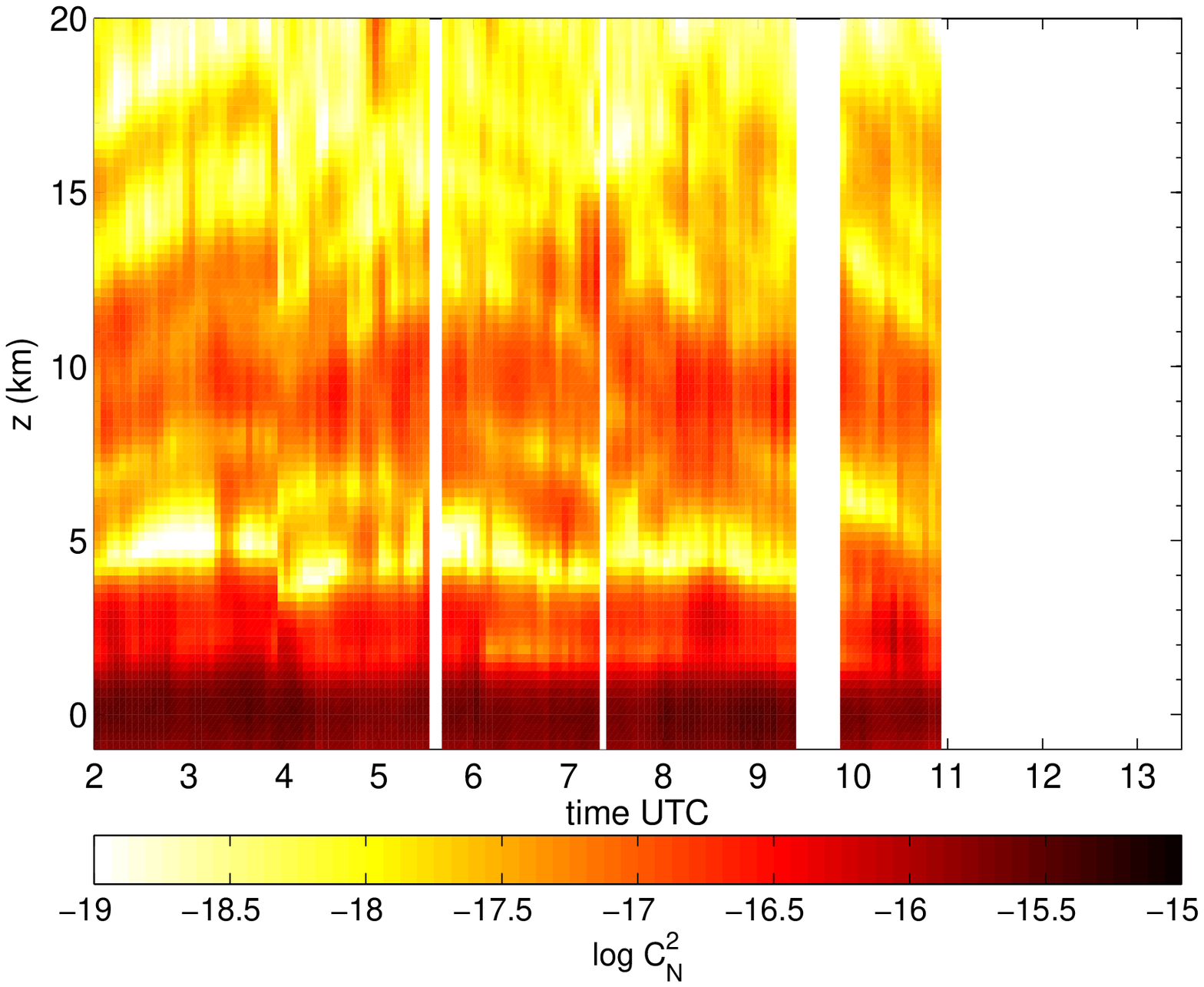}
\caption{Left: $\cn2$ temporal evolution predicted by the Meso-NH model on 1 March 2008 from 00 to 12 {\sc utc}. The simulated $\cn2$-profiles extend on 20 km. The temporal sampling is of one $\cn2$-profile each 2 min. Right: $\cn2$-profiles measured by the Generalized Scidar during the same night.}
\label{temp_evol}
\end{figure*}

The model is initialized and forced every 6 hours at the synoptic hours (00:00, 06:00, 12:00 {\sc utc}) with analyses from the the ECMWF (European Centre for Medium-Range Weather Forecasts). 
The model runs for 12 hours, the first two hours of every simulation are rejected because the model is still adapting to the orography. The output from the remaining 10 hours (19:00 - 05:00 {\sc lt})\footnote{LT = Local Time.} is used for the characterization of the optical turbulence at Mt Graham. The Astro-Meso-NH model provides the vertical profile of the $\cn2$ for every two minutes at the grid point located at the astronomical Observatory. Fig.~\ref{temp_evol} (left) shows an example of the temporal evolution of the $\cn2$ extended on 20 km obtained with the Astro-Meso-NH package. Fig.~\ref{temp_evol} (right) shows the correspondent temporal evolution of the measured $\cn2$-profiles obtained with the Generalized Scidar \citep{Mas10}. In these figures we can appreciate the characteristics of the simulated $\cn2$-profiles. Most of the turbulence layers at the different heights of the atmosphere appear well reconstructed by the model\footnote{Near the ground the shape of the $\cn2$ is necessarily different because the Generalized Scidar produces the typical bump due to the limited vertical resolution that spread the turbulent energy also under-ground.}. The spatial variability of the simulated turbulence is in general smoother than the measured one. This effect is more evident in the high part of the atmosphere. This is what we expect from the calculations of a parametrized parameter that is not explicitly resolved. Also the temporal variability of the simulated turbulence appears smoother with respect to the observed one. It is difficult to say that the value of a parametrized parameter can be predicted at a precise time t=t$_{0}$. For this reason so far we preferred to provide averaged estimates of the optical turbulence. We calculate the mean of the $\cn2$-profiles simulated of the whole night (with a sampling of the $\cn2$-profile every 2 minutes in the interval 19:00 - 05:00 {\sc lt}) and we associate the result to the mean of the observed $\cn2$-profiles obtained during the same night with the Generalized Scidar. We compare therefore typical measured and simulated $\cn2$-profiles. Hereafter we discuss the performance of the Meso-NH model under this assumption. This logic is the same used in all the previous studies done on the modelling of the optical turbulence with Meso-NH (\cite{Mas99b}, \cite{MasVB01}, \cite{MasGar01}, \cite{Mas02}, \cite{Mas04}, \cite{Mas06}, \cite{Las09}, \cite{Las10}).

\section{Model calibration}
\label{sec:calib}

\subsection{MJ01 method}
The method for the model calibration implemented in Meso-NH has been proposed by \cite{Mas01} (hereafter MJ01) and later on validated by \cite{Mas04}. It is based on the idea that the model depends on a free parameter (E$_{min}$ i.e. the minimum kinetic energy) that can be considered as a sort of background climatological noise. In the regions in which the dynamic turbulence is well developed the model rapidly forgets the E$_{min}$ value and the kinetic energy overcomes this value. The $\cn2$ is therefore not dependent on the E$_{min}$ in these regions. However, in stable regions, the authors proved that the $\cn2\propto E_{min}^{2/3}$. In each $\cn2$-profile it is possible to identify typically more than one vertical slab in which turbulence is in  stable regimes (MJ01-Fig.~1). If we change the value of E$_{min}$ it means that we assume that the atmosphere has more or less inertia in different vertical slabs and the thermodynamic instabilities require more or less energy to trigger turbulence in some regions of the atmosphere. As a consequence the threshold (the seed E$_{min}$) should be different in the respective vertical slabs. To identify the optimal values of E$_{min}$ the main idea in the MJ01 method was therefore to divide the whole atmosphere ($\sim$ 20 km) in a finite number (5-6) of vertical slabs and to optimize the value of  E$_{min}$ in each vertical slabs minimizing the differences between simulated  and measured $\cn2$-profiles i.e. minimizing the $\chi^{2}$ function for each night $m$:
\begin{equation}
\chi _{m,k}^{2} =\sum_{i=1}^{N_{k}}\left[ a_{m,k}\cdot
x_{m,i} -y_{m,i}\right] ^{2},
\end{equation}
where N$_{k}$ is the number of levels in the vertical slab k, $y_{m,i}$ is the average of the measured $\cn2$ sampled on N$_{k}$ levels for each night $m$, $x_{m,i}$ is the simulated $\cn2$ sampled on N$_{k}$ levels for each night $m$ and $a_{m,k}$ is the free coefficient that has to be fixed minimizing the function $\chi _{m,k}^{2}$. We refer the reader to \cite{Mas01} for details, we simply summarize here that, after an average of all the $a_{m,k}$ with respect to the number of nights $m$ we obtain a coefficient $a_{k}$ for each vertical slab. 
Knowing the value of $a_{k}$ for $k$ $=$ (1,K) (with K the number of vertical slabs), the kinetic energy is modified as:
\begin{equation}
E_{min,k}^{*}=E_{min}\cdot a_{k}^{3/2}\qquad k=1,K.
\label{emin}
\end{equation}
Once identified the optimized E$_{min}$, characterized by K steps, this is implemented in the model that is run again for each night. We highlight that the same new E$_{min}$ is used to simulate the new $\cn2$ for all the nights. This is what we call 'an output obtained with a calibrated model' and these are the results discussed in the next section, with respect to measurements.  

The MJ01 method reduces some systematic error and reconstructs a typical mean $\cn2$-profile better in agreement with measurements. The method has been validated statistically by \cite{Mas04} on a sample of 10 nights using measurements provided by different instruments and taken simultaneously.
It is worth to note that the method optimizes E$_{min}$ with respect to the mean $\cn2$-profiles simulated on the total sample used to calibrate the model. The calibration is based basically on the conservation of the turbulent energy (J=$\cn2\cdot\Delta$H) in each vertical slab. The differences between measurements and simulations calculated in each night, are in general, obviously larger than the difference between the statistical values (average in this case) as we will discuss later in the paper. The interest in increasing the statistical sample for the calibration is to verify how the reliability of the model changes increasing the number of nights. 

In this study the model calibration started with a selection of the sample of nights on which we calibrated the model. A brief digression is necessary here. In an ideal case, assuming to have a very rich sample of measurements (for example a year of measurements), one should calibrate the model using as many as possible different nights in order to approach a profile of E$_{min}$ characteristic of the site. Besides, once the model is calibrated, the ideal case should be to have an equally rich sample of independent measurements (and associated simulations) to investigate the performance of the model after calibration. It is also worth to highlight that the goal of the calibration is to reduce/eliminate a precise systematic error and therefore it is our interest to eliminate from the sample used for the calibration, all cases that corresponds to unusual results that might be associated to a failure of the model due to whatever reasons. These data, if introduced in the calibration sample, might bias the calibration. In other words, for the calibration a subjective selection is not only allowed but it is suggested without diminishing the confidence in the results.

In this study we can access to a sample of 43 nights. In spite of the great number of nights with respect to previous studies, it is still a too small number to be able to perform the strategy just described using two independent samples equally statistically rich. We therefore decided for an alternative solution that is a sort of compromise that maximize the progresses (answers to open questions) we can do in this field with this data-set. First, we are interested in limiting the sources of uncertainties in our sample of reference. The assumption done is necessarily to consider the measurements as a reference. In other words, we do the approximation that measured $\cn2$-profiles match with the true $\cn2$-profiles. For this reason we eliminated definitely from the original sample of 43 nights, two nights (March 2 and 3, 2008) because the relative shape of the mean of the $\cn2$-profile in each night is characterized by unusual features and these could affect the calibration process. Among the remaining 41 nights, we subtracted a sub-sample of seven nights ($\sim$ 17$\%$ of the total sample) in which the model provided unusual results that might let us think that for some reason the model did not work correctly\footnote{It is not important to discuss the causes for this potential discrepancies because we do not discard these nights from the total sample of nights but simply from the calibration sample because they might bias the calibration process. The typology of "unusual" results are for example, an overestimate of the turbulence near the ground. Whatever is the cause for this potential discrepancy (initialization data not representative of the atmospheric flow in a particular night or a specific failure of the model) we simply consider that this is what can happen in a real operational prediction.} independently from the algorithm used for the $\cn2$ parametrization. We therefore calibrate the model on the remaining sample of 34 nights that is the richest sample achievable with this data-set. We study therefore the statistical performances of the model calibration done on 34 nights separating the calibration aspect from the model score of success. After that, we quantify the model performances on the whole sample of 41 nights i.e. a sample including also the values of the seven nights that have not been considered in the calibration. As said before, this sample contains many cases that might be associated to bad model performances. This strategy permits us to do some steps ahead. We achieve, at the same time: (i) to optimize the calibration process, (ii) to discuss the model performance including an independent sample of nights not considered in the calibration and (iii) to investigate the model performance when it contains a realistic sample of failure cases. 

After the identification of the samples, we divided the atmosphere in six vertical slabs above (0 - 400 m), (400 m - 2 km), (2 - 7 km), (7 - 11 km), (11 - 13 km), (13 - 20 km) and we applied the model calibration for h $\ge$ 400 m. Below this height, we observed that the model does not depend any more on E$_{min}$ and the seed is quickly forgotten by the model during the simulation. As already explained in previous papers, the thickness and the number of the vertical slabs is arbitrary. We selected regions in which the turbulence seems to be characterized by similar trends.

\subsection{Variant to MJ01 method}
In this paper we tested if the calibration procedure for $E_{min}$ done on each model level instead of vertical slabs (as is the case of the MJ01 method) might produce substantially better results or not. The minimization of the $\chi^{2}$ function is done for each model level:

\begin{equation}
E_{min,i}^{*}=E_{min}\cdot a_{i}^{3/2}\qquad i=1,N.
\label{emin_b}
\end{equation}

where N is the total number of the model levels. This variant has been suggested by several colleagues in private communications after the publication of first results obtained with the MJ01 method \citep{Mas04, Mas06}. From a pure mathematical point of view such a method might in theory work better on the calibration sample (34 nights) because the mathematical fit has more constraints (a total number of measurements equal to the number of levels) on which to constrain the fit. However, from a physical point of view this method is characterized by a questionable physical assumption. It is indeed hard to justify it because it would be as to admit that the climatological noise $E_{min}$ is different at each model level but the values of $E_{min}$ are the same in each model levels for all the nights. This is the reason why we think that, at present, the original MJ01 version is the most solid approach from a physical point of view. However, in this paper, taking advantage of the rich statistical sample, we decided, independently from the arguments that might justify (or not) the MJ01 variant, to simply test this variant to verify, first, if the gain is effective and, second, to check if this gain is maintained with samples containing nights not included in the calibration sample (in our case the 41 nights sample). To simplify the discussion we call hereafter this method MJ01$^{*}$. 

\section{Results}
\label{sec:res}

\begin{figure}
\includegraphics[width=7cm]{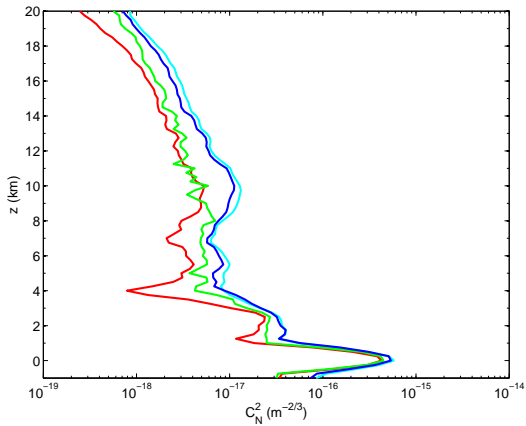}
\caption{Observed $\cn2$-profiles. Red Line: median of all the $\cn2$-profiles (15956) of the total sample of 41 nights. Green line: median of the 41 $\cn2$-profiles, each one is the average of the $\cn2$ of one night. Light blue line: mean of all the observed $\cn2$-profiles (15956) of the total sample of 41 nights. Dark blue line: mean of the 41 $\cn2$-profiles associated to each night and obtained averaging the $\cn2$ of one night.}
\label{cn2_mean_med}
\end{figure}

\begin{figure*}
\includegraphics[width=5.5cm]{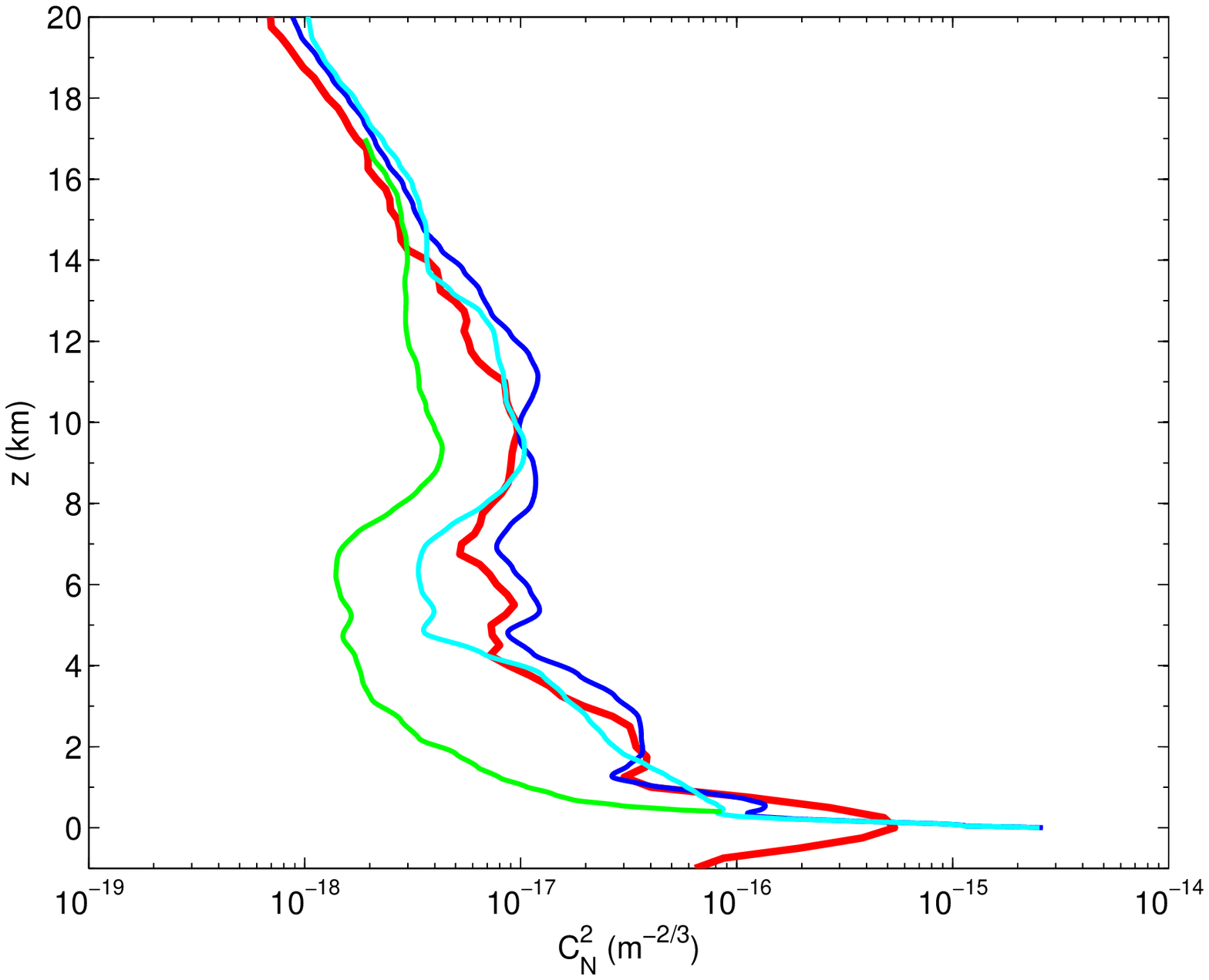}
\includegraphics[width=5.5cm]{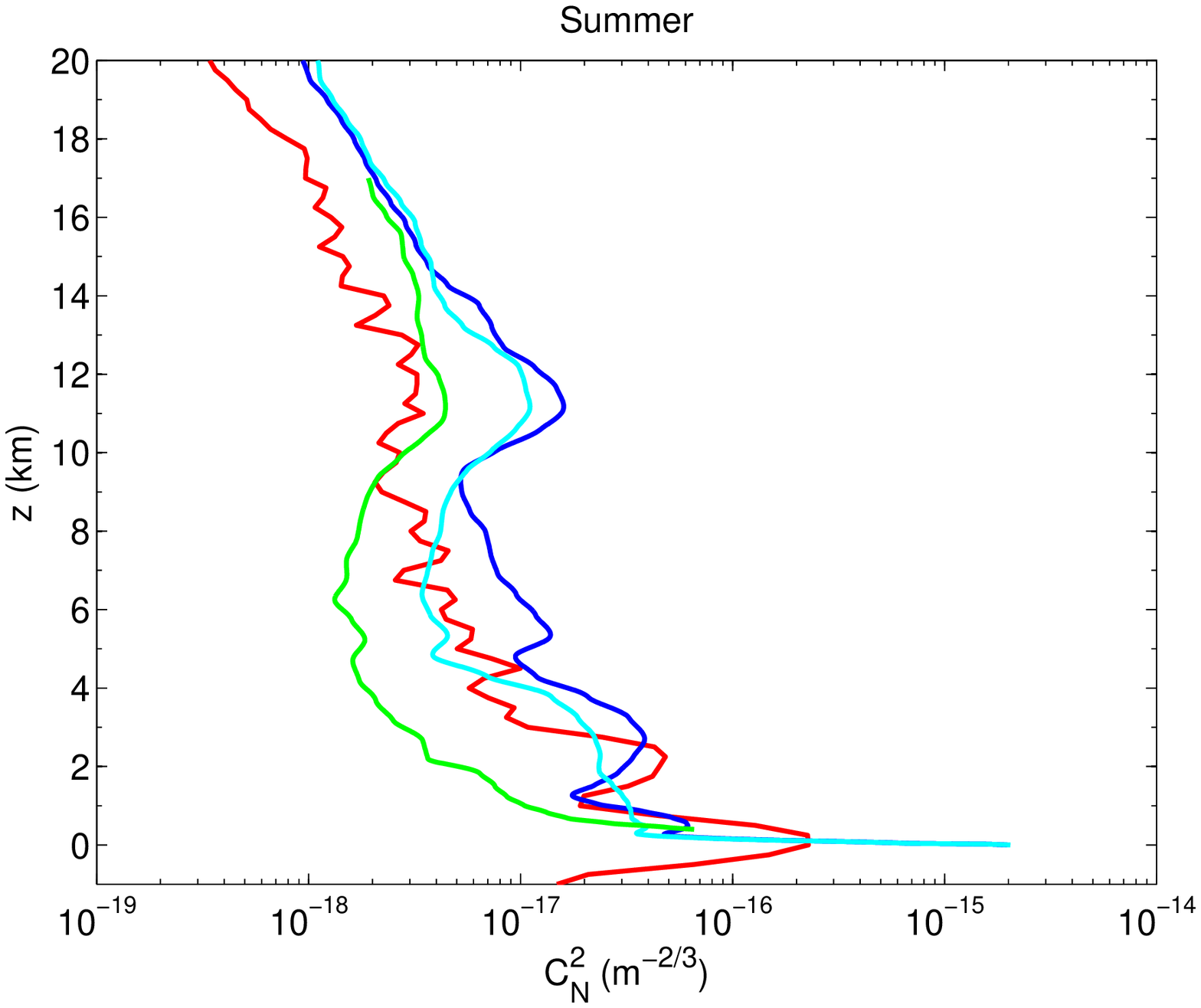}
\includegraphics[width=5.5cm]{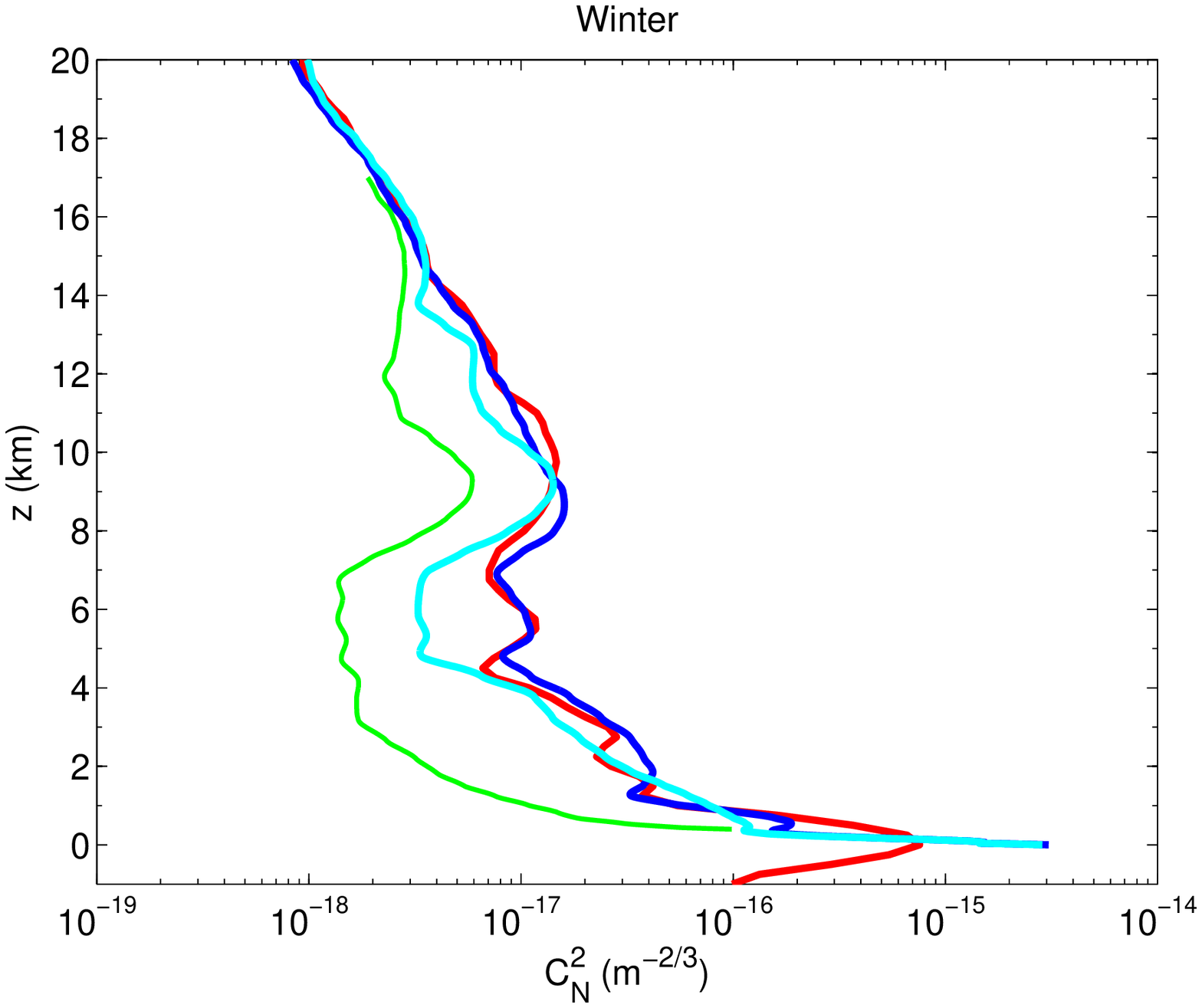}
\caption{The average $\cn2$-profiles measured by the Generalized Scidar (red line) and the Meso-NH model (light blue line: MJ01; dark blue line: MJ01$^{*}$) after calibration. Left: the average for the total sample of 34 nights. Centre: same for the summer (14 nights). Right: same for the winter (20 nights). The green line is plotted in the (400 m - 17 km) range and it shows the average $\cn2$-profile obtained by the model before the calibration.}
\label{cn2_34}
\end{figure*}

\begin{figure*} 
\includegraphics[width=5.5cm]{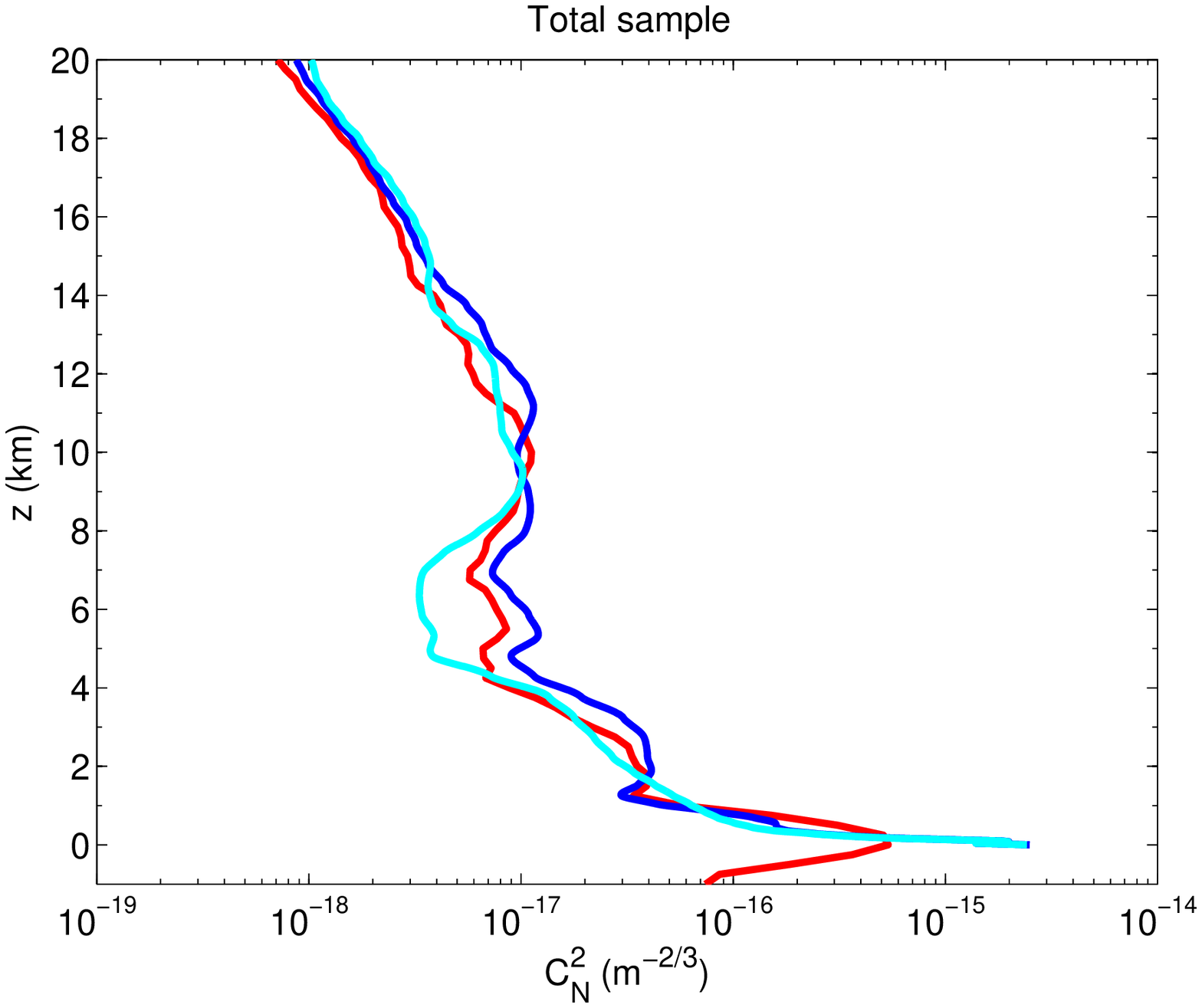}
\includegraphics[width=5.5cm]{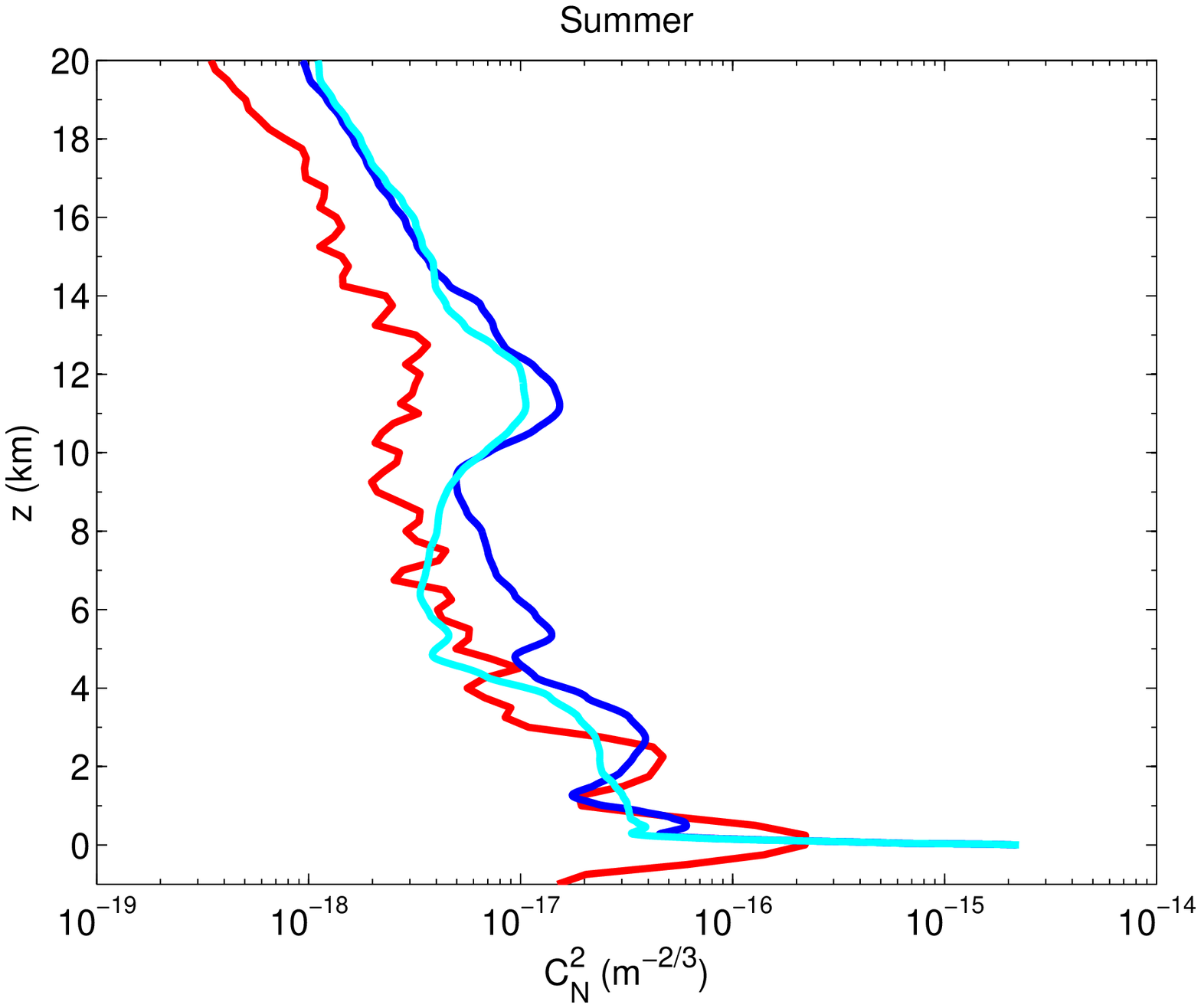}
\includegraphics[width=5.5cm]{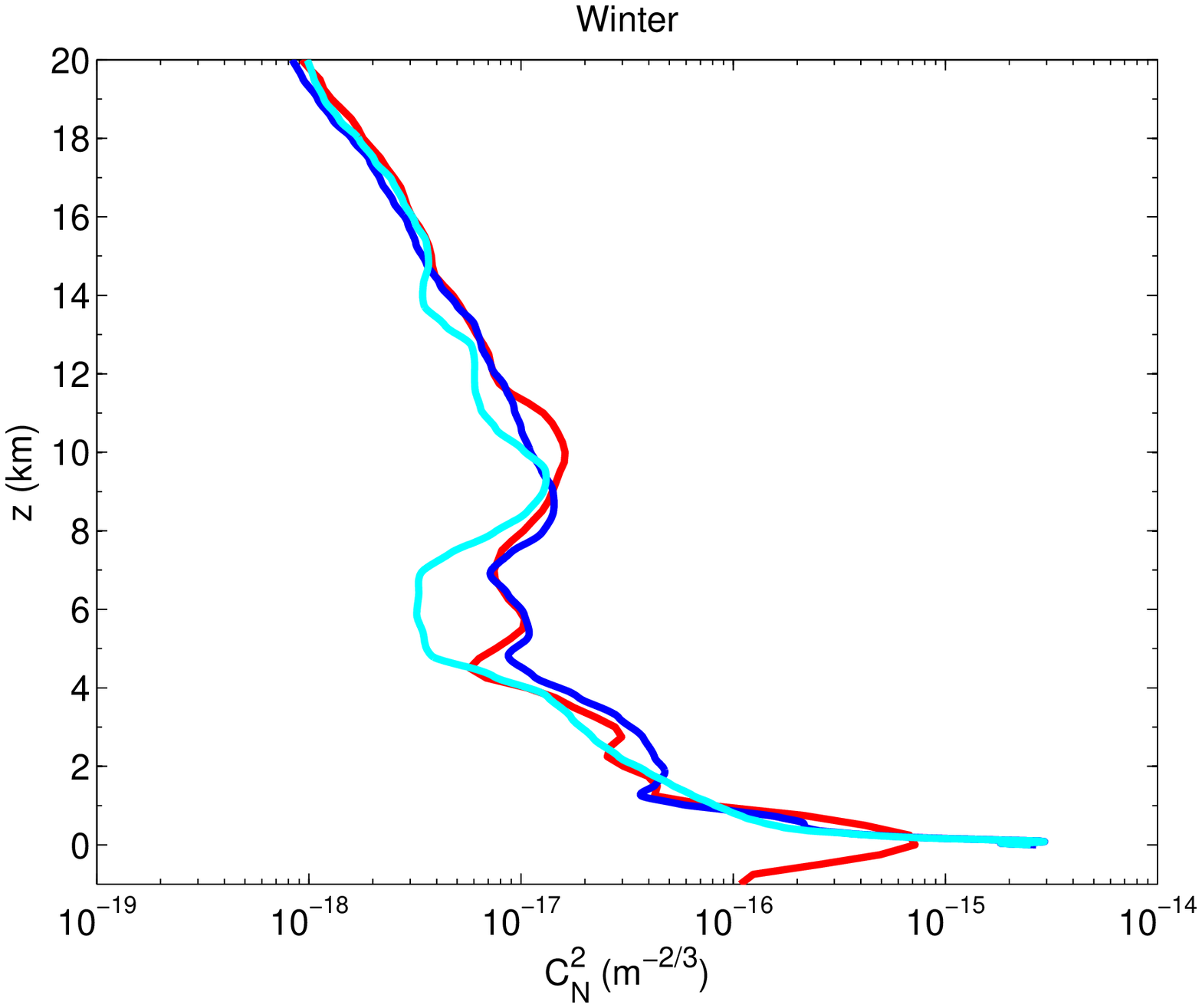}
\caption{The average $\cn2$-profiles measured by the Generalized Scidar (red line) and the Meso-NH model (light blue line: MJ01; dark blue line: MJ01$^{*}$) on the total sample of measured nights. Left: the average for the total sample of 41 nights. Centre: same for the summer (15 nights). Right: same for the winter (26 nights).}
\label{cn2_41}
\end{figure*}

\begin{figure} 
\includegraphics[width=7cm]{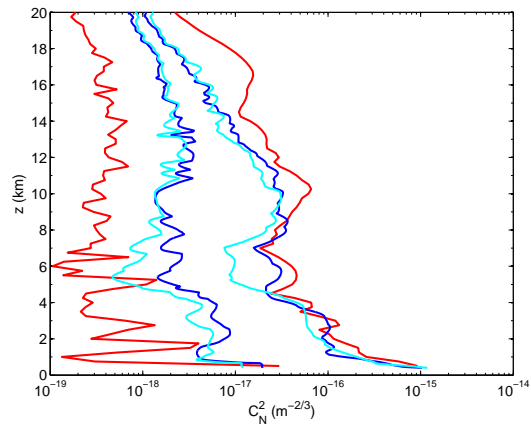}
\caption{Minimum and maximum values of the $\cn2$ observed and simulated in the sample of the 41 nights. Red lines refer to measurements from the GS. Light blue and dark blue lines refer respectively to the MJ01 and MJ01$^{*}$ models. Values of $\cn2$ are shown for h $\ge$ 400 m.}
\label{av_cn2_minmax}
\end{figure}

To discuss the results we divided the solar year in two periods: the summer (April - September) and the winter (October - March). We investigate the optical turbulence vertical distribution ($\cn2$-profiles) and the three major integrated astroclimatic parameters (seeing, isoplanatic angle and wavefront coherence time). For the $\cn2$ and the seeing  we show first the results obtained with the sample of 34 nights used for the calibration and, afterwards, the results obtained with total number of 41 nights. For the isoplanatic angle and the wavefront coherence time we treated directly the total sample of 41 nights. 

\begin{table*}
\caption{Calibration sample (34 nights): average seeing in the total atmosphere, boundary layer and free atmosphere. The dome seeing is removed from total seeing and the boundary layer seeing. The turbulence contribution provided by the first 20 m from the ground is excluded in the contribution of the Meso-NH model. The seeing is given in arcseconds using a wavelength of 500 nm.}
\begin{tabular}{lcccccccccccccc}
 & \multicolumn{3}{l}{Generalized Scidar}  & \multicolumn{5}{c}{Model MJ01}  & \multicolumn{5}{c}{Model MJ01$^{*}$} \\
   &$ \varepsilon_{tot}$ & $\varepsilon_{BL}$ & $\varepsilon_{FA}$   & & & $ \varepsilon_{tot}$ & $\varepsilon_{BL}$ & $\varepsilon_{FA}$ & & &$ \varepsilon_{tot}$ & $\varepsilon_{BL}$ & $\varepsilon_{FA}$\\
   \hline
Total  (34 nights) & 0.69 & 0.50 & 0.41 & & &  0.66 & 0.45 & 0.41 &  & &  0.74 & 0.49 & 0.48 \\
Summer  (14 nights) & 0.46 & 0.26 & 0.34 & & & 0.52 & 0.29 & 0.37 & & & 0.60 & 0.32 & 0.47 \\
Winter  (20 nights) & 0.85 & 0.66 & 0.46 & & & 0.76 & 0.56 & 0.43 & & & 0.84 & 0.60 & 0.49\\
\end{tabular}
\label{34_see}
\end{table*}


\begin{table*}
\caption{Relative error (measured in \%) calculated for the calibration sample and the total sample.}
\begin{tabular}{lrrrrrrrrrrrrrrrrrr}
\multicolumn{9}{c}{Calibration sample: 34 nights} & \multicolumn{10}{c}{Total sample: 41 nights} \\
\\
\multicolumn{4}{r}{Model MJ01} & \multicolumn{5}{r}{Model MJ01$^{*}$} & \multicolumn{5}{r}{Model MJ01} & \multicolumn{5}{r}{Model MJ01$^{*}$} \\
   & $\varepsilon_{tot}$ & $\varepsilon_{BL}$ & $\varepsilon_{FA}$ & & & $\varepsilon_{tot}$ & $\varepsilon_{BL}$ & $\varepsilon_{FA}$& &&$\varepsilon_{tot}$ & $\varepsilon_{BL}$ & $\varepsilon_{FA}$& & & $\varepsilon_{tot}$ & $\varepsilon_{BL}$ & $\varepsilon_{FA}$\\
\hline
Total   & 3.6 & 9.2 & 0.4 & & & 8.2 & 2.1 & 17.6 & & & 7.2&  9.0 & 3.2 & & & 18.0 & 15.4 & 13.2\\
Summer  & 13.5 & 11.6 & 13.0 & & & 31.9 & 21.1 & 36.8 & & & 13.8 &10.3  & 13.9 & & & 32.3 & 19.5  & 37.8\\
Winter  & 10.1 & 14.9 & 6.3 & & &0.7 & 8.4 & 7.6 & & & 5.1 & 8.7 & 10.4 & & & 13.6 & 14.4& 2.9\\
\end{tabular}
\label{err_rel}
\end{table*}


\begin{table*}
\caption{Total sample (41 nights): average seeing in the total atmosphere, boundary layer and free atmosphere. Same as Table~\ref{34_see}.}
\begin{tabular}{lcccccccccccccc}
 & \multicolumn{3}{l}{Generalized Scidar}& \multicolumn{5}{c}{Model MJ01}  & \multicolumn{5}{c}{Model MJ01$^{*}$} \\
 \\
   &$ \varepsilon_{tot}$ & $\varepsilon_{BL}$ & $\varepsilon_{FA}$ & & & $ \varepsilon_{tot}$ & $\varepsilon_{BL}$ & $\varepsilon_{FA}$ & & &$ \varepsilon_{tot}$ & $\varepsilon_{BL}$ & $\varepsilon_{FA}$ \\
   \hline
Total  (41 nights) & 0.71 & 0.51 & 0.43 & & & 0.76 & 0.56 & 0.41 & & & 0.84 & 0.59 & 0.48\\
Summer  (15 nights) & 0.45 & 0.26 & 0.34 & & &0.51 & 0.28 & 0.38 & & &  0.60 & 0.31 & 0.46\\
Winter  (26 nights) & 0.87 & 0.67 & 0.48 & & & 0.91 & 0.73 & 0.43 & & & 0.98 & 0.76 & 0.49\\
\end{tabular}
\label{41_see}
\end{table*}

When simulated versus measured $\cn2$-profiles are treated it is important to define how to statistically analyse the data. This is not univocal. As an example, Fig.~\ref{cn2_mean_med} shows, for a pedagogic approach, four $\cn2$-profiles calculated from the sample of observed $\cn2$-profiles: (i) the median of all the individual $\cn2$-profiles of the whole sample of 41 nights (15956 profiles) (red line); (ii) the median of the 41 $\cn2$-profiles associated to each night. Each $\cn2$-profile is obtained averaging the $\cn2$ of a night (green line); (iii) the mean of all the individual $\cn2$-profiles (15956 profiles) of the whole sample of 41 nights (light blue line); (iv) the average of the 41 $\cn2$-profiles associated to each night. Each $\cn2$-profile is obtained averaging the $\cn2$ of a night (dark blue line). The difference between these $\cn2$-profiles tells us that, depending on how we treat statistically the data we obtain different results. In studies on site characterization done with measurements taken with monitors approach (i) is frequently used. Our context, however, is different.

As we said, it is hard to predict a $\cn2$ at a precise time (t=t$^{*}$) of the night and, at present, to investigate quantitatively the performances of the model our goal is to associate the average of the $\cn2$-profiles simulated during one night with the average of the measured $\cn2$-profiles of the same night. For this reason we are forced to eliminate the approach (i) and (iii). Moreover the calibration is based basically on the fitting of the average of M $\cn2$-profiles (M is the number of nights). We selected therefore the criterion (iv) and we considered the 'mean' as a statistical operator to quantify the model performances of the $\cn2$. This is the same approach used in \cite{Mas04}. For homogeneity we considered the same criterion  with the integrated astroclimatic parameters (seeing, isoplanatic angle and wavefront coherence time).

\subsection{$\cn2$-profiles: optical turbulence vertical distribution}
\label{sec:cn2}

Figure~\ref{cn2_34} shows the average of the $\cn2$-profiles measured with the Generalized Scidar and simulated with the calibrated model with the methods MJ01 and MJ01$^{*}$. In the same picture is reported also the average $\cn2$ before the calibration as well as those calculated in the two seasons: summer and winter. Figure~\ref{cn2_41} shows the same for the 41 nights. 

The agreement between the morphology (shape) of the averaged $\cn2$-profiles measured by the Generalized Scidar (red lines) and calculated with the Meso-NH (light blue lines: MJ01; dark blue lines: MJ01$^{*}$) is very good when looking at the average of all the 34 nights (Fig.~\ref{cn2_34}-left). The model can reconstruct all the major typical features of the $\cn2$-profile such as the position (at $\sim$ 10 km) and shape of the secondary peak. Between 4 and 8 km above the ground the MJ01$^{*}$ method seems slightly better than the MJ01 method. The vertical distribution of the optical turbulence in the winter (Fig.~\ref{cn2_34} - right) is also very well described by the Meso-NH model. However in summer (Fig.~\ref{cn2_34} - centre) the model seems to overestimate the $\cn2$ in the free atmosphere. However, the MJ01 method is better than the MJ01$^{*}$ method. The MJ01 method is indeed well correlated to measurements up to around 8 km from the ground while the MJ01$^{*}$ method overestimates the $\cn2$ starting from 4 km from the ground. The {\it '$\alpha$ effect'} \citep{Mas06,Mas10} is well reconstructed by the model too: in summer the secondary peak shifts to a higher altitude as shown by the measurements. At the same time the strength of the secondary peak slightly decreases as observed in measurements. No major differences are appreciated in the shape of the averaged $\cn2$-profiles between the calibration case (34 nights) and the whole sample (41 nights). Therefore the inclusion in the sample of some more nights does not seem to produce a major impact on the morphology (shape) of the averaged $\cn2$-profile. Which is the cause of the model overestimate of the $\cn2$ in the free atmosphere in summer? Looking at Fig.~\ref{av_cn2_minmax} we note that, in the free atmosphere, the model variation between the minimum and the maximum values of $\cn2$ is substantially smaller than what observed with measurements. The higher the altitude, the higher is the model inertia and, as a consequence, the model has more difficulties in reconstructing the extreme values (minimum and maximum) in the high part of the atmosphere. One could expect therefore a slight underestimate of the model in winter and a slight overestimate in summer. The fact that a model overestimate is observed only in summer can be due to the fact that, in this season the number of nights for the 34 and 41 nights samples are, respectively, 14 and 15 nights while in winter we have 20 and 26 nights respectively. It is therefore possible that the winter data got a more important weight in the calibration. Also in summer we got all nights with extremely weak turbulence in the free atmosphere. With a richer and more homogeneous statistical sample we should therefore expect a decreasing in strength of the discrepancy and a more symmetric discrepancy with respect to the two seasons. We do not think that there is a specific problem for the summer period. Nevertheless it is a fact that the model variability needs to be improved in the high part of the atmosphere to be able to well detect all the cases of very strong as well as very weak turbulence conditions. We are working at present on this topic.   

\subsection{Seeing: $\varepsilon$}
\label{sec:see}

The seeing depends on the $\cn2$ as Eq.\ref{epsi} in Appendix A. Figure~\ref{see_fig} shows the total seeing ($\varepsilon_{tot}$), the boundary layer ($\varepsilon_{BL}$) and the free atmosphere ($\varepsilon_{FA}$) simulated by the model (MJ01 and MJ01$^{*}$) plotted against the respective seeing observed by the GS for the sample of 34 nights after calibration. The boundary layer is defined as the first kilometre above the surface and the free atmosphere from the top of the boundary layer up to 20 km above the surface. All values of the seeing are calculated opportunely by subtracting in the measurements from the GS the dome seeing contribution and by subtracting in simulations the contribution of first 20 m of the $\cn2$-profiles (the equivalent of the telescope height). In Figure~\ref{see_fig} (bottom-right of each panel) is reported the correlation coefficient calculated for the total sample of 41 nights. Table~\ref{34_see} and Table~\ref{err_rel}-left report the values of the average seeing and the relative error for the calibration sample (34 nights). Table~\ref{41_see} and Table~\ref{err_rel}-right reports the same for the total sample of 41 nights. The correlation coefficients are indicated for the MJ01 and MJ01$^{*}$ cases in each panel of Fig.~\ref{see_fig}. The total seeing and the seeing in the boundary layer show a good correlation (c.c. = 0.78-0.82) with measurements in the calibration sample with no major differences between the MJ01 and MJ01$^{*}$ cases. It is very encouraging that the turbulence near the ground, that represents most of the turbulence developed in the atmosphere, is well predicted by the model. The correlation decreases for the seeing in the free atmosphere. The explanation for such an effect is that, in the free atmosphere, visibly the model inertia is still too high and the parameters predicted by the model varies in a smaller range than what the measurements show as discussed in Section \ref{sec:cn2}. 

\begin{figure*}
\includegraphics[width=6cm,angle=-90]{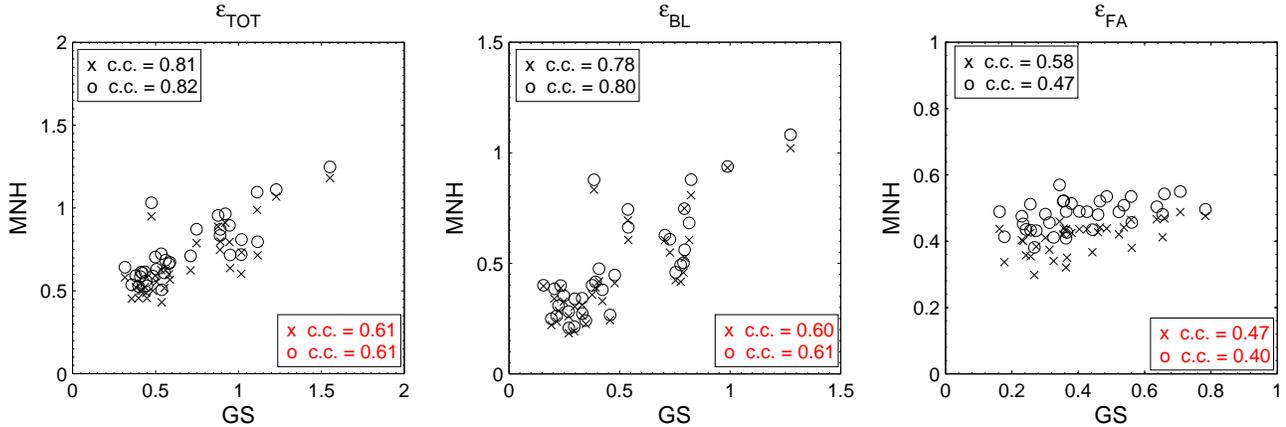}
\caption{Values of the seeing from the simulations (MJ01 xs, MJ01$^{*}$ circles) plotted against the measurements in the calibration sample (34 nights). The left hand panel shows the total seeing, the middle panel the seeing in the boundary layer, the right hand panel the free atmosphere seeing. The dome seeing has been removed from the Generalized Scidar data and the seeing in the first 20 m (correspondent to the height of the telescope) has been removed from the Meso-NH simulations. In bottom-left corner of each panel the correlation coefficient calculated for MJ01 and MJ01$^{*}$  for the calibration sample is reported. In bottom-right corner (red) the c.c. calculated for the total sample of 41 nights.
\label{see_fig}}
\end{figure*}

\begin{figure*}
\includegraphics[width=5.5cm]{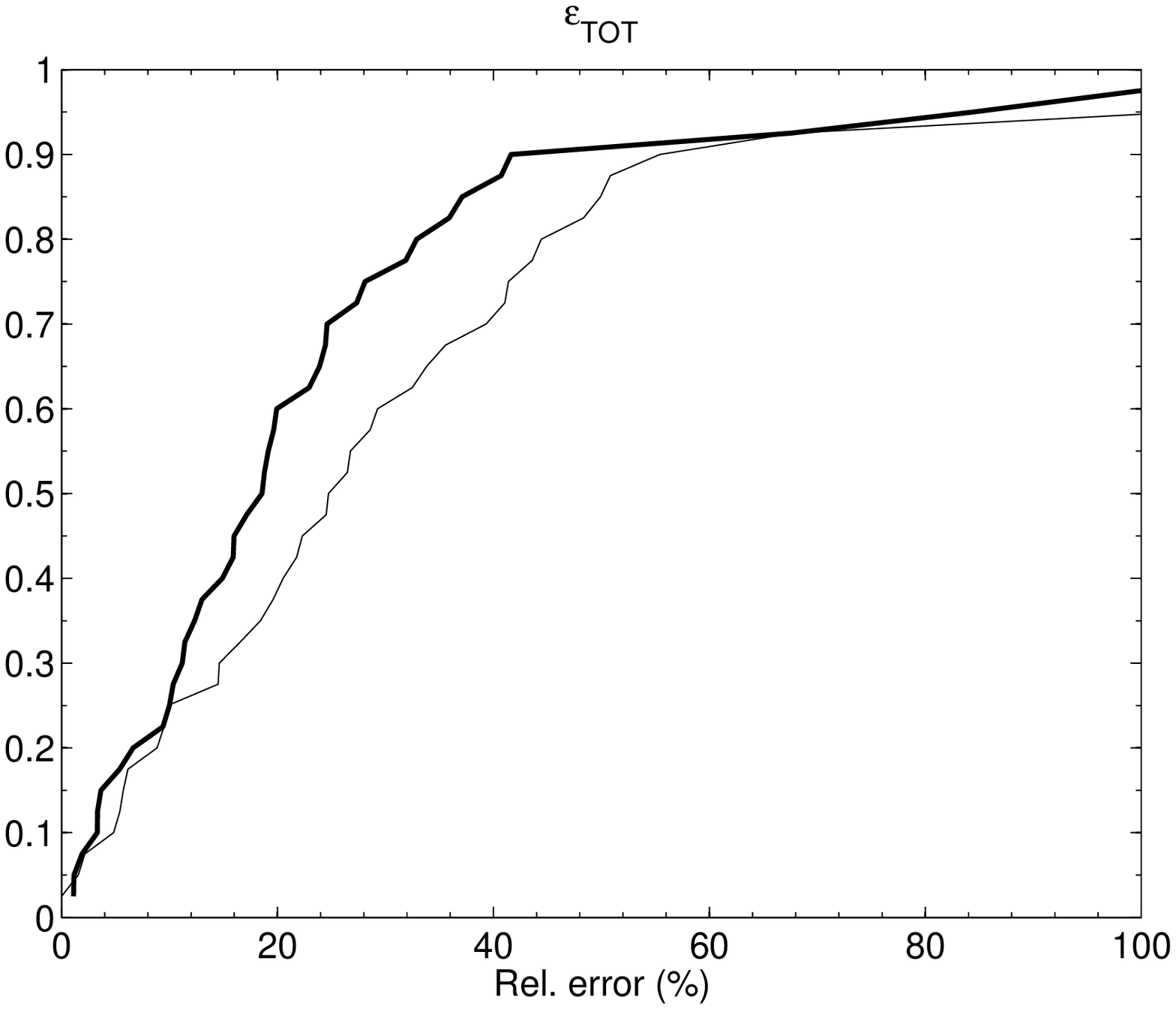}
\includegraphics[width=5.5cm]{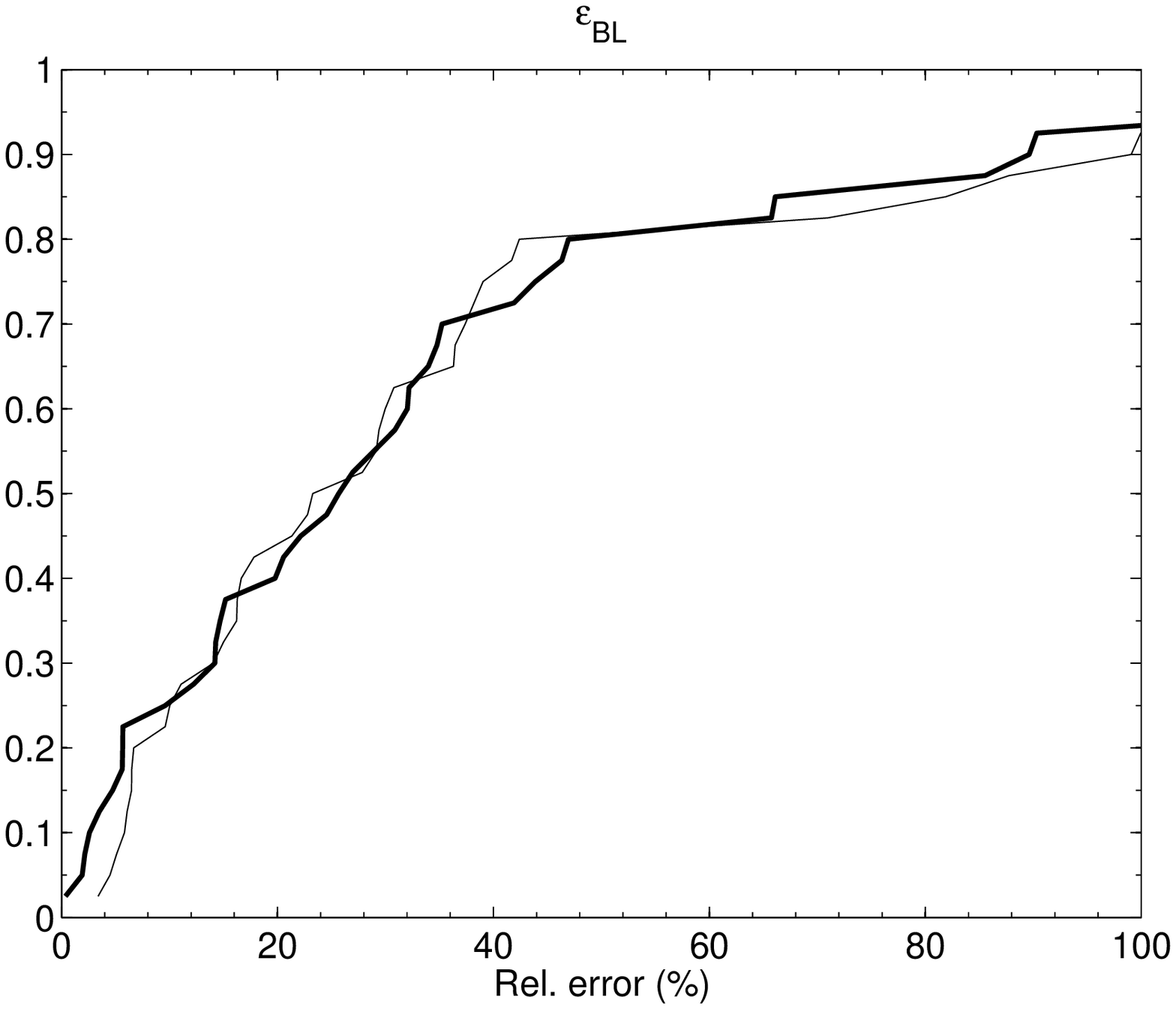}
\includegraphics[width=5.5cm]{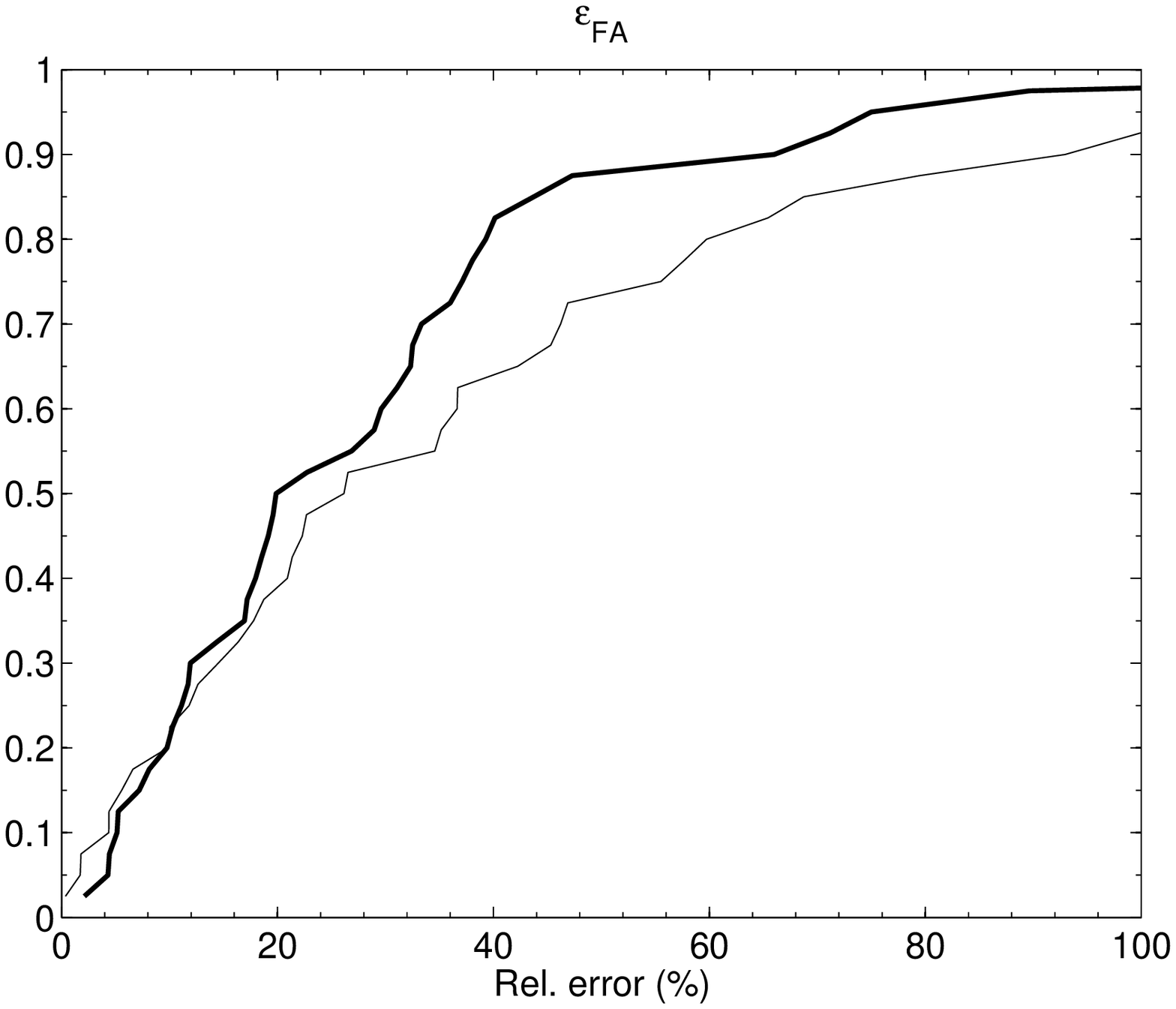}
\caption{Cumulative distribution of the relative errors of each single night of the total seeing ($\varepsilon_{TOT}$), the seeing in the boundary layer ($\varepsilon_{BL}$) and the seeing in the free atmosphere ($\varepsilon_{FA}$) for the total sample of 41 nights. Bold line: MJ01 method. Thin line: MJ01$^{*}$ method. }
\label{see_sing}
\end{figure*}

The model has at present some problems in identifying the best (the minimum observed $\varepsilon_{FA}$) and the worst (the maximum observed $\varepsilon_{FA}$) conditions in this part of the atmosphere. Looking at Fig.~\ref{cn2_34} it appears clear that the problem concerns the $\cn2$ at h $>$ 10 km. However, we note that the model (with both methods, MJ01 and MJ01$^{*}$) is able to reconstruct a weaker $\varepsilon_{FA}$ in summer than in winter as we expect and also as is observed (Table~\ref{34_see}). 
On the sample of calibration no substantial and systematic differences are noted on results obtained with the MJ01 and MJ01$^{*}$ methods. The very encouraging result is that the relative error for the seeing in all the three regions of the atmosphere (Table \ref{err_rel}-left) is very good even in the free atmosphere. 
Results are equally very good if we look at the individual seasons. 
The important conclusion is that the relative error for the seeing on the whole sample is within 9.2\% for the best method (MJ01) and within 17.6\% if we consider also MJ01$^{*}$ method\footnote{This indicates that increasing the number of nights the calibration attain more accurate results. \cite{Mas04} on a sample of 10 nights obtained a relative error on the average of the order of 20\%.}. If we consider the two sub-samples (summer and winter) the relative error decreases slightly still maintains good performances. 
In conclusion we observe relative errors remarkably good in spite of a modest correlation in the free atmosphere. We also note that the correlation is very good for seeing smaller than 1 arcsec and deviations are more typical when there is strong turbulence. In conclusion we can say that the method MJ01 confirms to be effective and it improves the model reliability reducing some systematic effects. 

If we look at the total sample of 41 nights (Table~\ref{41_see} and Table~\ref{err_rel}-right), including nights not used for the calibration we observe that the correlation decreases for $\varepsilon_{tot}$ and $\varepsilon_{BL}$ (from 0.82 to 0.61). For $\varepsilon_{FA}$, the correlation decreases in a less consistent way and the MJ01 method seems better than the MJ01$^{*}$ method in this region. 
Most of the new nights included in this sample belong to the winter period and the reason for the decrease of the correlation is that the model reconstructs a seeing in the boundary layer that is statistically too strong.
In this case, the MJ01 method is visibly better than the MJ01$^{*}$ method in both cases: the total sample and the summer and winter periods. Method MJ01 achieves a remarkably good result with a relative error within 9.0\% for the total sample and a maximum relative error of 13.9\% if we look at the sub-samples (summer and winter). The MJ01 method is also better than MJ01$^{*}$ method in reconstructing the weakest and the strongest values in the two season in the partial $\varepsilon_{FA}$ and $\varepsilon_{BL}$. 
Also in this case the model reconstructs a $\varepsilon_{FA}$ that varies in a smaller range with respect to what observed. Equally to the calibration sample, we can however observe that the model can reconstruct a weaker seeing in summer than in winter as expected.

Finally, to study the performance of the model night by night in a compact way, we have calculated the cumulative distribution of the relative errors  with respect to measurements for the $\varepsilon_{tot}$, $\varepsilon_{BL}$ and $\varepsilon_{FA}$ obtained with the MJ01 method and the MJ01$^{*}$ method (Fig.~\ref{see_sing}). We note that the relative error in the three regions of the atmosphere is absolutely remarkably good. The median values (50\% of times) of the relative error is within 26.3\% in the three regions of the atmosphere for the MJ01 method, which is visibly better than the MJ01$^{*}$ method in this case. This means that, even if the correlation for the $\varepsilon_{FA}$ is not as good as for $\varepsilon_{tot}$ and $\varepsilon_{BL}$, the relative error night by night is maintained within 21\%. The biggest relative error is obtained in the boundary layer. Even if this is the region in which the correlation is the best one, the impact of an error of the model can be more important than an error produced in the free atmosphere.

In conclusion we can state that, on the total sample of nights, the MJ01 method provides globally better results than the MJ01$^{*}$ method. This means that the MJ01 method, which is based on a minor number of constraints than the MJ01$^{*}$ method, has a better performance probably because it is based on more robust physical assumptions. A larger number of constraints can be effective on the calibration sample but, when an independent sample of nights is taken into account, this theoretical advantage seems to lose its effect. 


\subsection{Isoplanatic angle: $\theta_{0}$}
\label{sec:iso}

\begin{figure}
\begin{center}
\includegraphics[width=5.5cm]{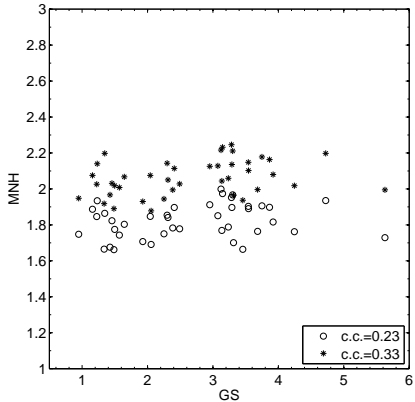}
\caption{Isoplanatic angle ($\theta_{0}$) from the simulations (MJ01 stars, MJ01$^{*}$ circles) plotted against the measurements related to the sample of 41 nights. }
\end{center}
\label{41_iso}
\end{figure}

Figure~\ref{41_iso} shows the simulated versus the measured isoplanatic angle $\theta_{0}$ for the sample of 41 nights. $\theta_{0}$ depends on the $\cn2$ as Eq.\ref{theta0} in Appendix A. The values reconstructed by the model span a much smaller range (1.7-2.2) arcsec than what has been observed (0.9-5.6) arcsec. The average of the values obtained for all the nights is in good agreement with measurements. The average of the simulations is equal to 2.06 arcsec for MJ01 (1.82 arcsec for MJ01$^{*}$) versus 2.65 arcsec measured by the GS. The difference between simulations and measurements is due to a small off-set in the model calibration. As we have already discussed in the previous sections, the calibration sample is slightly biased in the high part of the atmosphere by the fact that measurements are not completely homogeneously distributed in the two seasons and in the summer we have many nights with extremely weak turbulence in the high part of the atmosphere. Looking at Fig.~\ref{cn2_41}-left we can appreciate a generally very good agreement between measurements and simulations but  a very weak overestimate by the model is visible in the $\cn2$ above 10 km. $\theta_{0}$ is very sensitive even to small variation of the $\cn2$ in this region since it scales with $\cn2(h)\cdot h^{5/3}$. Therefore a very weak difference in the $\cn2$ in the high part of the atmosphere can produce sub-arcsecond differences in the isoplanatic angle. The problem of the small off-set in the average should disappear if one uses for the model calibration a very rich and homogeneous statistical sample of nights (one year for example). However, the failure of the model in reconstructing the larger variability of $\theta_{0}$ quantified by measurements is not due to the calibration but to the fact that, at present, the model shows a larger inertia in the high part of the atmosphere as discussed in Section~\ref{sec:cn2}. This produces a more modest correlation between simulations and measurements for $\theta_{0}$. As already said we are at present working on this issue to improve the model variability. 

\begin{figure}
\begin{center}
\includegraphics[width=5.5cm]{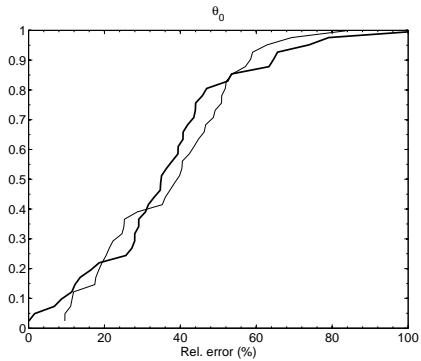}
\caption{Cumulative distribution of the relative errors of each single night of the isoplanatic angle ($\theta_{0}$) for the total sample of 41 nights. Bold line: MJ01 method. Thin line: MJ01$^{*}$ method.}
\end{center}
\label{iso_sing}
\end{figure}

Figure~\ref{iso_sing} shows the cumulative distribution of the relative error (night by night) for $\theta_{0}$ obtained with both methods (MJ01 and MJ01$^{*}$). The median value of relative error for $\theta_{0}$ is slightly larger (35.1 $\%$ for the best method, MJ01) than what is observed for the seeing described in the previous section. 

\subsection{Wavefront coherence time: $\tau_{0}$}
\label{sec:tau}

\begin{figure*}
\includegraphics[width=5.5cm]{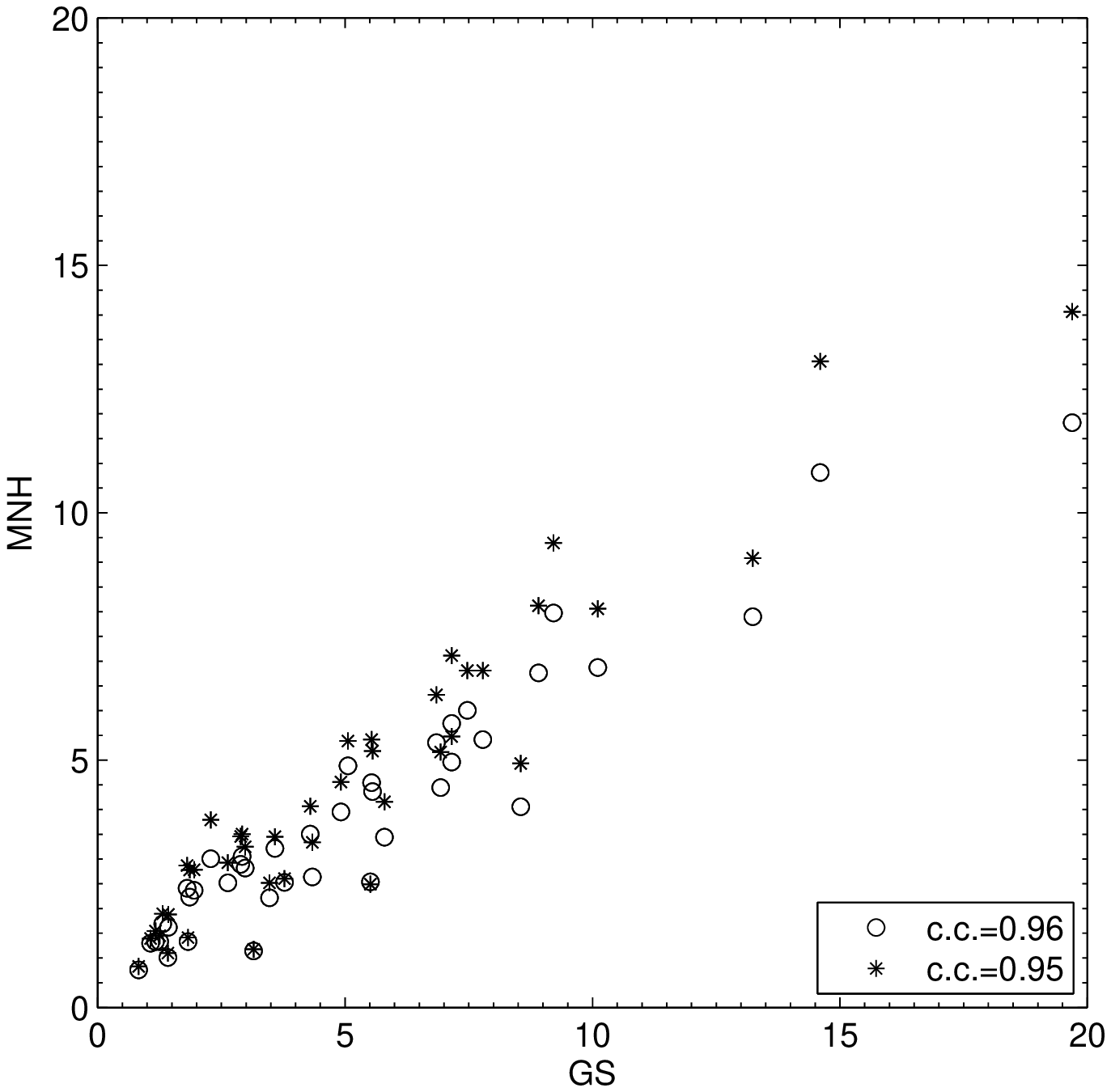}
\includegraphics[width=5.5cm]{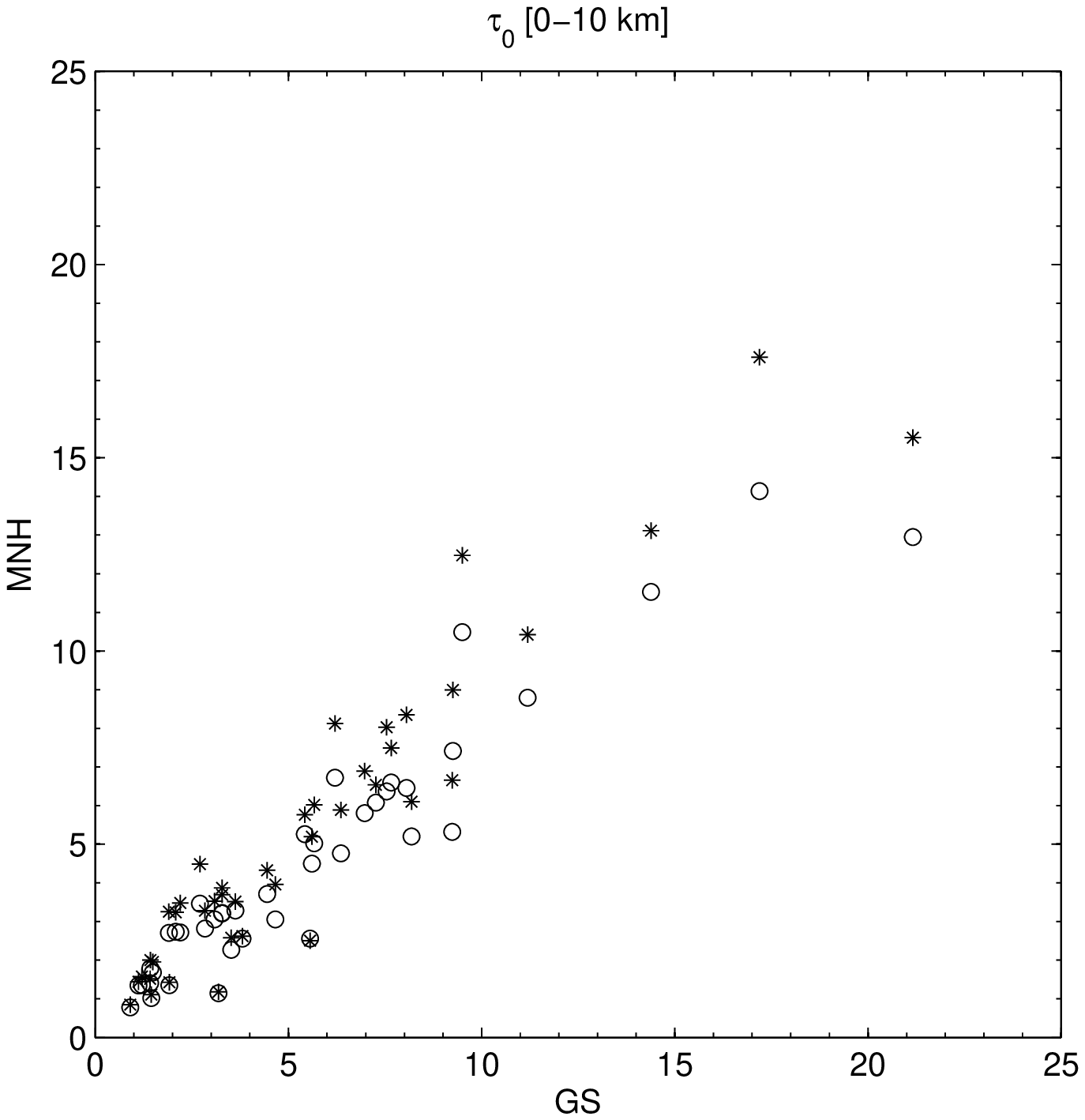}
\includegraphics[width=5.5cm]{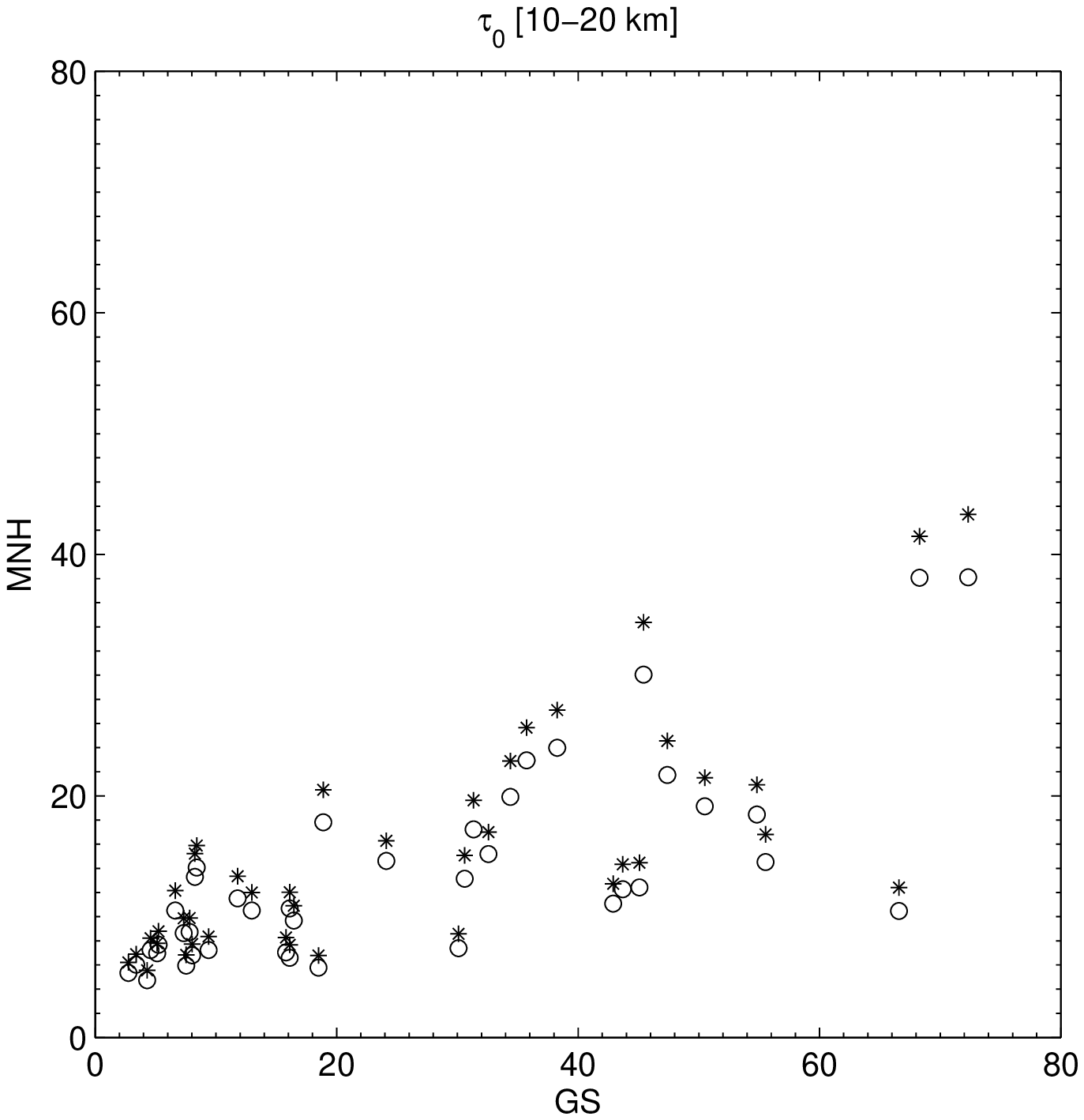}
\caption{Wavefront coherence time ($\tau_0$) from the simulations (MJ01 stars, MJ01$^{*}$ circles) plotted against the measurements related to the sample of 41 nights. 
\label{41_tau}}
\end{figure*}

$\tau_{0}$ depends on the $\cn2$ and the wind speed vertical profiles as Eq.\ref{tau0} in Appendix A. 
How do we calculate $\tau_{0}$ for measurements and simulations? For the wind speed profiles we have showed in previous studies \citep{Egner07, Mas10} that the best solution is to consider a composite profile: the wind speed profiles retrieved from the ECMWF analyses (i.e. analyses from General Circulation Model) in the nearest grid point ($\sim$12 km away) for $h$ above 1 km and the wind speed as retrieved from the GS for $h$ below 1 km. This is necessary because the analyses are not sampled with enough high horizontal resolution and they are not reliable near the ground. However, in a more recent study \citep{Hag10} it has been proved that the wind speed profiles provided by the Meso-NH model at the summit of Mt Graham are very well correlated to the wind speed estimated by the ECMWF analyses\footnote{The authors also showed that the wind speed is uniform on a horizontal scale of some tens of kilometres for h $>$ 1 km. The wind speed from analyses close to Mt Graham are in agreement with radiosoundings launched from Tucson International airport ($\sim$120 km from Mt Graham) and this guarantees the reliability of analyses.} above 1 km and are well correlated to measurements taken with a GS and an anemometer near the ground. To calculate the $\tau_{0}$ we considered therefore the wind speed reconstructed by Meso-NH above the summit. Due to the fact that for both, measured and simulated $\tau_{0}$, we used the wind speed profiles as retrieved from the Meso-NH, when comparing the $\tau_{0}$  we are basically comparing the effects of the simulated and measured $\cn2$ on the simulated and measured $\tau_{0}$. We consider the average of the wind speed profiles during each night and we calculate the $\tau_{0}$ for each night. Figure~\ref{41_tau}-left shows the simulated versus the measured wavefront coherence time ($\tau_{0}$) for the sample of 41 nights. The data-set is well distributed along a straight line showing a very good correlation (0.95 for MJ01 and 0.96 for MJ01$^{*}$). Up to 10 msec the values are very well correlated. A very small bias is evident for the extremely good values of $\tau_{0}$ ($\ge$ 10 msec). 

To verify which part of the $\cn2$ (low or high atmosphere) mainly affects $\tau_{0}$ in Fig.~\ref{41_tau}-left, we calculated a $\tau_{0}$ in the partial 0-10 km and 10-20 km ranges (Fig.~\ref{41_tau}-centre and Fig.~\ref{41_tau}-right). We observe that while the simulated and measured values are well correlated in the 0-10 km range, the model underestimates the $\tau_{0}$ in the 10-20 km range. However the contribution coming from the 0-10 km region has a much more important affect on the calculation of $\tau_{0}$ on the whole atmosphere than the contribution coming from the 10-20 km region. This does that $\tau_{0}$ on the whole 0-20 km is very well reconstructed by the model globally. This is a confirmation that, also for this parameter, the small bias of the $\cn2$ in the high part of the atmosphere produces some effect on $\tau_{0}$. However, differently from $\theta_{0}$, this effect is almost negligible (Fig.~\ref{41_tau}-left). 

\begin{figure}
\begin{center}
\includegraphics[width=5.5cm]{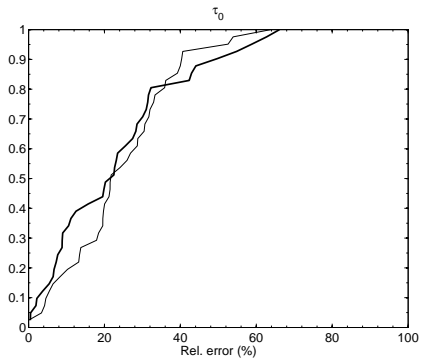}
\caption{Cumulative distribution of the relative errors of each single night of the wavefront coherence time ($\tau_{0}$) for the 41 nights. Bold line: MJ01 method. Thin line: MJ01$^{*}$ method.}
\end{center}
\label{tau_sing_nights}
\end{figure}

To study the performance of the model night by night, Fig.~\ref{tau_sing_nights} shows the cumulative distribution of the relative error for $\tau_{0}$, simulated with both the MJ01 and MJ01$^{*}$ methods, and calculated, taking the measurements as a reference. The median of the relative error is very good for $\tau_{0}$, as good as for the seeing, 22.5 $\%$ for the MJ01 method and 21.8 $\%$ for the MJ01$^{*}$ method. 

\section{Conclusions}
\label{sec:concl}
In this study we discussed the abilities of the Meso-NH model in simulating the optical turbulence above Mt Graham, site of the Large Binocular Telescope. Simulated $\cn2$-profiles are compared to a large statistical sample of measured $\cn2$-profiles related to 41 nights obtained with a Generalized Scidar. This large sample of measurements allowed us to study the performances of the model calibration in great detail and to investigate the performance of the model in different seasons discussing statistically how well the model reconstruct the $\cn2$, the seeing in different regions of the atmosphere, the isoplanatic angle and the wavefront coherence time. 

Two different methods of model calibrations have been investigated. The calibration is done on a sample of 34 nights and the model performances are discussed on the whole sample of 41 nights. 

The most important results obtained are:\newline
(i) We have proved that the model calibration definitely improves the model performance. The morphology of the vertical distribution of the optical turbulence (average of all $\cn2$ of the
34 nights) matches very well with the corresponding measured $\cn2$-profile. If we look at the model behaviour in different seasons, we show evidence of a model overestimate of $\cn2$ in the high part of
the atmosphere (h $>$ 8 km) in summer. This bias is highly probable because the sample of the investigated nights is not completely uniformly distributed between summer and winter. With a reasonably richer sample (of the order of 1 yr) this small bias can be corrected. This evidence proves the necessity of a very rich statistical sample to carry out an efficient model calibration. 
\newline
(ii) We observed that the model basically always reconstructs the most important features of the shape of the measured $\cn2$-profile but it still has some difficulties in reconstructing the very extreme values (very good and very bad turbulence conditions) in the free atmosphere. We are working at present on improving these model performances. \newline
(iii) For the model calibration, the total and boundary layer seeing reconstructed by the Meso-NH model are well correlated with measured values with a correlation coefficient (c.c.) of the order of 0.78-0.82. The seeing in the free atmosphere is more weakly correlated to measurements (c.c. $\sim$0.58). The relative errors are however extremely good in all the three regions (total seeing, in the boundary layer and free atmosphere). The best method MJ01 provide a relative error within 9.2$\%$ on the total sample and within 14.9\% if we consider the two seasons. These percentages remain basically the same if we consider the whole sample of 41 nights. \newline
(iv) When we consider the total sample of 41 nights the correlation between simulations and measurements decreases slightly for $\varepsilon_{tot}$ and $\varepsilon_{BL}$ (c.c. $\sim$0.60), much less for $\varepsilon_{FA}$ (c.c. $\sim$0.47).\newline
(v) If we consider the cumulative distribution of the relative errors night by night (i.e. the typical conditions of the operational mode) we find extremely encouraging results. The median value of the relative errors is indeed extremely small with some small variations for the seeing in the three regions of the atmosphere but typical of the order of 20$\%$. Therefore even if we look at the most difficult conditions typical of the operational mode, the model maintains a reasonable good result. \newline
(vi) For the isoplanatic angle the relative error of the average is still very good ($\sim$ 18\%) but the median of the relative errors calculated night by night is 35.1\%. It is the parameter with the poorest performances between $\varepsilon$, $\theta_{0}$ and $\tau_{0}$. This is not surprising because it is the parameter that, more than others, is very sensitive to the turbulence in the high part of the atmosphere.
 \newline
(vii) The wavefront coherence time reconstructed by the model shows a very good correlation with measurements (c.c. $\sim$ 0.95). We proved that the small bias in the high atmosphere on the $\cn2$ produces a negligible effect on $\tau_{0}$. The first 0-10 km represent the most important contributions to the final $\tau_{0}$ value. The cumulative distribution of the relative error calculated night by night is still very good with a median value equal to 22.5\%. \newline
(viii) The MJ01 method to calibrate the model confirms to be a solid method and it shows a general better performance of the model than the variant MJ01$^{*}$, which consists in fitting measurements and simulations on each model level instead of vertical slabs of a few kilometers. It is worth to note that the calibration of the high atmosphere is more delicate because a small discrepancy from the measured $\cn2$-profile can produce a large discrepancy on some astroclimatic parameters (such as $\theta_{0}$).\newline
The general conclusions are that the Meso-NH model is able to describe the optical turbulence distribution above Mt Graham showing very good performances with small relative errors with respect to measurements for most of the astroclimatic parameters. The most urgent problem to be solved is the improvement of the model ability in reconstructing the very weak and very strong turbulent conditions in the high atmosphere, i.e.~the ability in reconstructing isoplanatic angles in better agreement with measurements. Finally we have put in evidence the importance to have a very rich statistical sample of $\cn2$-profiles to efficiently calibrate the model. In the future we would like to access an even much richer samples of measurements done preferably with a GS in order to be able to consider a completely independent sample for the model validation. In the future it would also be interesting to test the model performance when it is initialized with forecasts from the ECMWF instead of the operational analyses in order to use the Meso-NH model as a part of a system to forecast the optical turbulence at Mt Graham allowing the implementation of a flexible-scheduling management of the LBT. 

\appendix

\section{Integrated astroclimatic parameters as a function of the $\cn2$.}

The seeing ($\varepsilon$), the isoplanatic angle ($\theta_{0}$), the wavefront coherence time ($\tau_{0}$) are defined as:

\begin{equation}
r_0  = \left[ {0.423 \cdot \left( {\frac{{2\pi }}
{\lambda }} \right)^2  \cdot \int\limits_0^\infty  {C_N^2 (h)dh} } \right]^{ - 3/5} 
\end{equation}

\begin{equation}
\varepsilon  = 0.98\frac{\lambda }
{{r_0 }}
\label{epsi}
\end{equation}

\begin{equation}
\theta _0  = 0.057 \cdot \lambda ^{6/5}  \cdot \left[ {\int\limits_0^\infty  {h^{5/3} C_N^2 (h)dh} } \right]^{ - 3/5} 
\label{theta0}
\end{equation}

\begin{equation}
V_0  = \left[ {\frac{{\int\limits_0^\infty  {V(h)^{5/3} C_N^2 (h)dh} }}
{{\int\limits_0^\infty  {C_N^2 (h)dh} }}} \right]^{3/5} 
\end{equation}

\begin{equation}
\tau _0  = 0.31\frac{{r_0 }}
{{V_0 }}
\label{tau0}
\end{equation}

\section*{Acknowledgements}
ECMWF products are extracted from the catalogue MARS, http://www.ecmwf.int. Access to these data was authorized by the Meteorologic Service of the Italian Air Force. We thank the VATT and the LBT staff for the support offered during the GS runs at Mt Graham.
This study has been funded by the Marie Curie Excellence Grant (FOROT) - MEXT-CT-2005-023878.


\appendix

\label{lastpage}
\end{document}